\documentclass[aps,prd,showpacs,superscriptaddress,nofootinbib,floatfix,10pt]{revtex4-2} 
\usepackage[utf8]{inputenc}
\usepackage{color}
\usepackage{amsfonts,amsmath,amssymb,leftindex}
\usepackage{graphicx}
\usepackage{epstopdf}
\usepackage{longtable,booktabs}
\usepackage{bm}
\usepackage[colorlinks=true,linkcolor=blue,citecolor=blue]{hyperref}
\usepackage{geometry}
\usepackage{url}
\usepackage{epstopdf}
\usepackage{mathtools}
\usepackage{subfigure}
\usepackage{multirow}
\usepackage{diagbox}
\usepackage{afterpage}
\usepackage{placeins}
\usepackage{float}
\usepackage{mathrsfs}
\usepackage{setspace}
\usepackage{orcidlink}
\allowdisplaybreaks[1]
\graphicspath{{figs/}}

\renewcommand{\arraystretch}{1.5}

\newcommand{\itp}{\affiliation{CAS Key Laboratory of Theoretical Physics, Institute of Theoretical Physics,\\ Chinese Academy of Sciences, Beijing 100190, China}}

\newcommand{\ucas}{\affiliation{School of Physical Sciences, University of Chinese Academy of Sciences, Beijing 100049, China}}

\newcommand{\peng}{\affiliation{Peng Huanwu Collaborative Center for Research and Education, Beihang University, Beijing 100191, China}}

\newcommand{\scnt}{\affiliation{Southern Center for Nuclear-Science Theory (SCNT), Institute of Modern Physics,\\ Chinese Academy of Sciences, Huizhou 516000, China}}

\newcommand{\qfnu}{\affiliation{College of Physics and Engineering, Qufu Normal University, Qufu 273165, China}}

\newcommand{\chep}{\affiliation{School of Physics and Center of High Energy Physics, Peking University, Beijing 100871, China}}

\begin{document}

\title{Coupled-channel analysis of the near-threshold \texorpdfstring{$e^+e^-\to N\bar{N}$}{e+e-toNNbar} cross sections}
\author{Zhao-Sai Jia\orcidlink{0000-0002-7133-189X}}\thanks{These authors contributed equally to this work.}\qfnu \itp

\author{Zhen-Hua Zhang
\orcidlink{0000-0001-6072-5378}}\thanks{These authors contributed equally to this work.}
\chep \itp

\author{Feng-Kun Guo\orcidlink{0000-0002-2919-2064}} \email{fkguo@itp.ac.cn}
\itp \ucas \peng \scnt

\author{Gang Li\orcidlink{0000-0002-5227-8296}}\email{gli@qfnu.edu.cn} \qfnu

\begin{abstract}
 
The possible existence of nucleon-antinucleon bound states has been studied for decades.
We investigate the $e^+e^-\to p\bar{p}$ and $e^+e^-\to n\bar{n}$ cross sections in the nonrelativistic effective field theory framework. The proton-antiproton and neutron-antineutron coupled-channel final state interactions are considered and found responsible for near-threshold enhancements. Both the proton-neutron mass difference and the Coulomb interaction between $p$ and $\bar{p}$ are considered, and the $N\bar{N}$ strong interactions are taken into account through a short-distance optical potential. 
By fitting the low energy constants in the amplitudes to the data for the near-threshold $e^+e^-\to N\bar{N}$ cross sections from the BESIII and SND collaborations, a $N\bar{N}$ quasibound state is found just above the $p\bar{p}$ threshold, and another $N\bar{N}$ pole is found on the unphysical Riemann sheet, farther away from the threshold. The constructed coupled-channel amplitude with Coulomb effects also offers a framework that can be used directly in experimental analyses on fine structures near the $N\bar{N}$ thresholds.
\end{abstract}

\maketitle

\section{Introduction}
The interest in the nucleon-antinucleon bound states has been lasting for decades~\cite{Klempt:2002ap}. They are analogs of the deuteron but with a vanishing baryon number.
One prominent candidate of a $p\bar{p}$ bound state is the $X(1835)$ observed in the $\eta^\prime\pi^+\pi^-$ invariant mass distribution by the BES collaboration~\cite{BES:2005ega} in the $J/\psi\to \gamma \eta^\prime \pi^+\pi^-$ process in 2005. 
It could correspond to the near-threshold enhancement in the $p\bar{p}$ final state spectrum of the $J/\psi\to \gamma p\bar{p}$ decay observed by the BES collaboration in 2003~\cite{BES:2003aic}. 
The $X(1835)$ and the $p\bar{p}$ near-threshold enhancement were later confirmed by the BESIII~\cite{BESIII:2010vwa,BESIII:2010gmv} and CLEO~\cite{CLEO:2010fre} experiments with higher statistics, and their quantum numbers were determined to be $J^{PC}=0^{-+}$~\cite{BESIII:2011aa,BESIII:2015xco}. In 2013, a narrower structure $X(1840)$ with $J^{PC}=0^{-+}$ was reported by the BESIII collaboration in the $6\pi$ spectrum of the $J/\pi\to \gamma 3(\pi^+\pi^-)$ decay~\cite{BESIII:2013sbm}, and $X(1880)$ in the intermediate vicinity of the $p\bar p$ threshold was reported using the full dataset of the BESIII experiment in 2024~\cite{BESIII:2023vvr}. 
The higher precision data for the $X(1835)$ and $X(1840)$ demonstrated line shapes distorted from the Breit-Wigner distribution, which could be attributed to the $p\bar{p}$ threshold effect or the interference between two resonances~\cite{BESIII:2016fbr,BESIII:2023vvr}. 
On the other hand, no near-threshold enhancement was observed in the $J / \psi \rightarrow \omega p \bar{p}$ process~\cite{BESIII:2013lac}.
Many theoretical efforts have been made to explain these anomalous near-threshold structures. Besides the $p\bar{p}$ bound state~\cite{Ding:2005gh,Wang:2006sna,Dedonder:2009bk,Liu:2009vm,Niu:2024cfn,Ortega:2024zjx} or the final state interaction (FSI) effect~\cite{Zou:2003zn, Sibirtsev:2004id, Chen:2008ee, Chen:2010an,Chen:2011yu,Kang:2015yka}, the resonances were also interpreted as, e.g., pseudoscalar glueballs~\cite{Li:2005vd,Kochelev:2005vd,Hao:2005hu,Kochelev:2005tu,Gui:2019dtm} and radial excitation states of the $\eta^\prime$ meson~\cite{Huang:2005bc,Yu:2011ta,Wang:2020due} (for reviews, see~\cite{Liu:2016yfr,Ma:2024gsw}), while no consensus has been reached so far. Proton-antiproton
% $p\bar{p}$ 
near-threshold enhancements have also been observed in other processes, e.g., in the $p\bar{p}$ spectrum in the $B$ decays~\cite{Belle:2002bro,Belle:2002fay,Belle:2007oni} and $\psi(2S)$ decays~\cite{BESIII:2010vwa,CLEO:2010fre,BESIII:2011aa}, as well as in the $e^+e^-\to p\bar{p}$ process~\cite{BaBar:2005pon,BaBar:2013ves,CMD-3:2015fvi}. In particular, the enhancement in $e^+e^-\to p\bar{p}$ is of particular interest as it could contain the contribution of possible $N\bar{N}$ ($N=p,n$) vector bound state(s) with $J^{PC}=1^{--}$. 

The enhancement in $e^+e^-\to p\bar{p}$ has been known for a long time. In 1994, the PS170 collaboration~\cite{Bardin:1994am} first 
reported the steep energy dependence of the $p\bar{p}\to e^+e^-$ cross section near the $p\bar{p}$ threshold. The 
near-threshold enhancement was later confirmed by the FENICE collaboration~\cite{Antonelli:1994kq} in the cross section for $e^+e^-\to p\bar{p}$ with large uncertainties, and the first measurement of the $e^+e^-\to n\bar{n}$ cross section~\cite{Antonelli:1998fv} was also made by the FENICE group. Since then, there have been updated measurements of the $e^+e^-\to N\bar{N}$ cross sections by the BABAR~\cite{BaBar:2005pon,BaBar:2013ves}, CMD-3~\cite{CMD-3:2015fvi,CMD-3:2018kql}, BESIII~\cite{BESIII:2015axk,BESIII:2019tgo,BESIII:2019hdp,BESIII:2021rqk,BESIII:2021tbq}, and SND~\cite{Achasov:2014ncd,SND:2022wdb,Achasov:2024pbk} collaborations, providing more precise data near the $N\bar{N}$ thresholds. 
Particularly, the CMD-3 collaboration measured the $e^+e^-\to p\bar{p}$ cross section and found that the steep rise happened within about 1~MeV above the $p\bar p$ threshold~\cite{CMD-3:2018kql}. 
It has been shown that the near-threshold enhancement in the $e^+e^-\to p\bar{p}$ cross section can be explained by the FSI between the $p$ and $\bar{p}$ in a series of theoretical studies~\cite{Haidenbauer:2006dm,Dmitriev:2007zz,Haidenbauer:2014kja,Haidenbauer:2015yka,Dmitriev:2015qyt,Milstein:2018orb,Milstein:2022tfx}, where the $N\bar{N}$ interaction potential is given by models such as the Paris model~\cite{El-Bennich:2008ytt,Dmitriev:2007zz}, Nijmegen model~\cite{Zhou:2012ui}, and J\"ulich model~\cite{Hippchen:1991rr,Mull:1991rs,Mull:1994gz,Kang:2013uia,Haidenbauer:2014kja}, or the chiral effective field theory~\cite{Chen:2011yu,Kang:2013uia,Kang:2013uia,Dai:2017ont,Xiao:2024jmu}. All of these potentials are optical potentials~\cite{Carbonell:2023onq}, whose imaginary parts account for contributions from the strong annihilation channels of the $N\bar{N}$ far below the $N\bar{N}$ thresholds, e.g., $n\pi$ $(n\geq2)$, and these different models give similar results. 
However, most of these theoretical works ignored the Coulomb interaction between $p$ and $\bar{p}$ and the proton-neutron mass difference, as such effects should be sizable only within a few MeV above the $p\bar p$ threshold while the data studied in the literature concentrated on a higher energy region. 
The first coupled-channel model including the Coulomb potential in the position space was given by Ref.~\cite{Milstein:2018orb} and improved in Ref.~\cite{Milstein:2022tfx} to analyze the CMD-3 data with fine structures, and the model produced a clear cusp at the $n\bar{n}$ threshold. An isoscalar pole about $4-20$ MeV above the $p\bar{p}$ threshold in the $\leftindex^{(2S+1)}L_J= \leftindex^3 {S}_1$ (or $\leftindex^3 S_1$-$\leftindex^3 D_1$ mixing) partial wave was found in the $N\bar{N}$ scattering matrix in many theoretical works~\cite{Kang:2013uia,Haidenbauer:2014kja,Dmitriev:2015qyt,Haidenbauer:2015yka,Milstein:2018orb,Xiao:2024jmu}.
Such a pole is located on the physical Riemann sheet (RS) of the complex energy plane and is referred to as unstable or quasistable bound state in the literature. It is also compatible with the complex $p\bar{p}$ scattering length in the $\leftindex^3 S_1$-$\leftindex^3 D_1$ channel extracted from the energy level shift of the antiprotonic hydrogen (the $\bar{p}H$ hadronic atom)~\cite{Gotta:1999vj,Gotta:2004rq,Carbonell:2023onq} via the Trueman relation~\cite{Trueman:1961zza}. It is pushed away from the real axis to the complex energy plane by  lower annihilation channels, and would become a bound state below the threshold when these annihilation channels are turned off~\cite{Dmitriev:2015qyt,Xiao:2024jmu}. 
In Ref.~\cite{Yan:2023nlb}, the experimental data for the $e^+e^- \to \bar{p} p $ and $n\bar{n}$ total cross sections above thresholds up to about 3~GeV have been reproduced by introducing three excited vector mesons using the vector-meson dominance model.
Recently, the SND collaboration~\cite{Achasov:2024pbk} has just released precise data for the $e^+e^-\to n\bar{n}$ near-threshold cross section, which offers great opportunities to study the possible near-threshold $N\bar{N}$ poles with $J^{PC}=1^{--}$ in detail.

In this work, we investigate the $e^+e^-\to p\bar{p}$ and $e^+e^-\to n\bar{n}$ near-threshold cross sections in the nonrelativistic effective theory (NREFT) framework. The $N\bar{N}$ FSIs are taken into account in the Watson-Migdal approach~\cite{Watson:1952ji,Migdal1955THEORYON}, including the $p\bar{p}$--$n\bar{n}$ coupled-channel dynamics and the Coulomb interactions between $p$ and $\bar{p}$. The leading order (LO) strong interaction between the $N$ and $\bar{N}$ is parametrized by a constant optical potential, and the corresponding low energy constants (LECs) are determined by a combined  fit to the newly updated data for $e^+e^-\to n\bar{n}$ near-threshold cross sections from the SND collaboration~\cite{Achasov:2024pbk} and the $e^+e^-\to p\bar{p}$ near-threshold cross section measured by the BESIII collaboration~\cite{BESIII:2021rqk}. 
The near-threshold $N\bar{N}$ pole positions can then be derived.

% The rest part of this paper 
This paper is organized as follows. In Sec.~\ref{Sec:Cross sections}, we derive expressions for the $e^+e^-\to N\bar{N}$ cross sections in the coupled-channel NREFT. In Sec.~\ref{Sec:Results}, the parameters are fixed by fitting to the experimental cross section data, and the pole positions of the $N\bar{N}$ coupled-channel scattering amplitudes in the near-threshold region are obtained with the fixed LECs. In Sec.~\ref{results with fixed a11}, we perform another fit with the $p\bar p\to p\bar p$ scattering length fixed to that extracted from the antiprotonic hydrogen, in order to check the consistency between our results and the atomic measurement. A brief summary is given in Sec.~\ref{Sec:Summary}.

\section{Cross sections in coupled-channel NREFT}\label{Sec:Cross sections}

In this section, we give the amplitudes and cross sections of $e^+e^- \to N\bar{N}$ near the $N\bar{N}$ thresholds in the $p\bar{p}$--$n\bar{n}$ coupled-channel NREFT. The $p\bar{p}$ and $n\bar{n}$ are referred to as the first and second channels, respectively. The two reactions $e^+e^-\to p\bar{p}$ and $e^+e^-\to n\bar{n}$ should proceed dominantly via the one-photon exchange~\cite{Haidenbauer:2006dm,Haidenbauer:2014kja,Haidenbauer:2015yka}, so the quantum numbers of the $N\bar{N}$ pair are fixed to be $J^{PC}=1^{--}$. The near-threshold energy dependence of the cross sections is mainly given by that of the $N\bar{N}$ FSI amplitudes, which contain contributions from both Coulomb and strong interactions.

The transition operator of the $N\bar{N}$ scattering including the Coulomb contribution can be expressed in the two-potential formalism as~\cite{Kong:1999sf}
\begin{align}
    \hat{T}(E)=\hat{T}_C(E)+\hat{G}_0(E)^{-1}\hat{G}_C(E)\hat{T}_{SC}(E)\hat{G}_C(E)\hat{G}_0(E)^{-1},
\end{align}
where $E=\sqrt{s}$ is the total initial energy in the $N\bar{N}$ center-of-mass (c.m.) frame; $\hat{T}_C$ and $\hat{T}_{SC}$ are the Coulomb and strong-Coulomb transition operators, respectively; $\hat{G}_0=1/(E-\hat{H}_0+i\varepsilon)$ is the free Green's resolvent with $\hat{H}_0$ the free Hamiltonian; and $\hat{G}_C=1/(E-\hat{H}_0-\hat{V}_C+i\varepsilon)$ the Coulomb Green's resolvent with $\hat{V}_C$ the Coulomb potential. 
The transition operators satisfy the Lippmann-Schwinger equations (LSEs)
\begin{align}
    \hat{T}_C&=\hat{V}_C+\hat{V}_C\hat{G}_0\hat{T}_C,\\
    \hat{T}_{SC}&= \hat{V}_S+\hat{V}_S\hat{G}_C\hat{T}_{SC},
\end{align}
where $\hat{V}_S$ is the strong potential between $N$ and $\bar{N}$.

\begin{figure}[tb]
    \centering
    \includegraphics[width=0.9\linewidth]{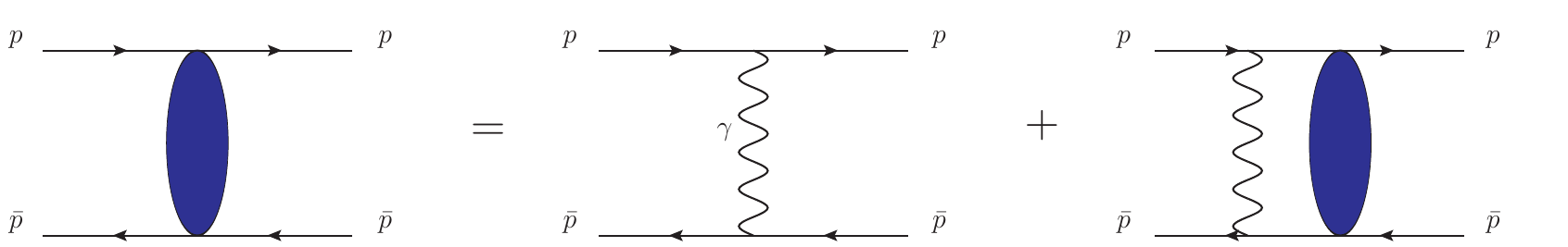}
    \caption{Resummation in the Coulomb $T$ matrix for the $p\bar{p}$ scattering. The wavy line represents a photon, and the blue blob represents resummation of the Coulomb photon exchanges.}
    \label{fig:Coulomb_resum}
\end{figure}
The $N\bar{N}$ system with $J^{PC}=1^{--}$ can only be in $\leftindex^3 S_1$ and $\leftindex^3 D_1$  partial waves~\cite{Haidenbauer:2006dm}. 
We mainly consider the $S$-wave contribution in this work, as the near-threshold $D$-wave amplitude is suppressed by $k_N^2$ comparing to the $S$-wave one, with $k_N$ the c.m. momentum of the nucleon $N$. 
The $S$-wave two-channel scattering $T$ matrix reads~\cite{Braaten:2017kci,Shi:2021hzm}
\begin{align}
    \bm{T}(E) = \begin{pmatrix}
 T_C(E) & 0 \\
 0 & 0
\end{pmatrix} 
+ \begin{pmatrix}
    W_C(E) & 0 \\
    0 & 1 \end{pmatrix} 
    \bm{T}_{SC}(E) 
    \begin{pmatrix}
    W_C(E) & 0 \\
    0 & 1 \end{pmatrix},
    \label{Eq_Tmatrix}
\end{align}
where
\begin{align}
    T_C(E) = \frac{i \pi}{ \mu_p k_p} \left( \frac{\Gamma(1-i x)}{\Gamma(1+i x)} - 1\right)
    \label{Eq_TC}
\end{align}
is the $p\bar{p}$ Coulomb scattering amplitude shown in Fig.~\ref{fig:Coulomb_resum} including infinite Coulomb photon exchanges, with $\mu_p=m_p/2$ the reduced mass of $p\bar{p}$, $k_p=\sqrt{2\mu_p(E-2m_p)}$ the c.m. momentum of the proton, $x=\alpha\mu_p/k_p$, and $\alpha=1/137$ the fine-structure constant; the $\bm{T}_{SC}(E)$ is the strong-Coulomb scattering matrix; and 
\begin{align}
    W_C(E)=\left(\frac{2 \pi x}{1-e^{-2 \pi x}} \frac{\Gamma(1-i x)}{\Gamma(1+i x)}\right)^{{1}/{2}}
    \label{Eq_WC}
\end{align}
includes the effect of resumming the Coulomb photon exchanges between the $p\bar{p}$ pair in the initial or final state.

Since we focus on the immediate vicinity of the $N\bar{N}$ near thresholds, it is reasonable to employ NREFT at LO, which has a constant strong interaction potential for the $S$ wave. The scattering states in the two channels can be expressed in terms of the isospin basis as 
\begin{align}
    \left| p\bar{p} \right \rangle=-\frac{1}{\sqrt{2}}\bigg(\left| N\bar{N}, I=0 \right \rangle+\left| N\bar{N}, I=1 \right \rangle\bigg),\nonumber\\
    \left| n\bar{n} \right \rangle=-\frac{1}{\sqrt{2}}\bigg(\left| N\bar{N}, I=0 \right \rangle-\left| N\bar{N}, I=1 \right \rangle\bigg),
    \label{Eq_I=01}
\end{align}
and the LO strong potential reads
\begin{align}
   \bm{V}_S
    =\frac{1}{2}\begin{pmatrix}
        C_{0\text{N}}+C_{1\text{N}}+\delta_{\text{em}} & C_{0\text{N}}-C_{1\text{N}} \\
        C_{0\text{N}}-C_{1\text{N}} & C_{0\text{N}}+C_{1\text{N}}
    \end{pmatrix}\equiv \begin{pmatrix}
        a_1 + i a_2 & b_1 + i b_2 \\
        b_1 + i b_2 & c_1 + i a_2
    \end{pmatrix},
    \label{Eq_VSCN}
\end{align}
where $C_{0\text{N}}$ and $C_{1\text{N}}$ are LECs for the isoscalar and isovector interactions, respectively, and $\delta_{\text{em}}$ represents electromagnetic and isospin breaking corrections on the $p\bar{p}$ strong interaction (see below)~\cite{Shi:2021hzm}. The $\bm{V}_S$ matrix elements are complex valued, as reparametrized in terms of real free parameters $a_{1,2}$, $b_{1,2}$, and $c_1$.
The imaginary parts of the LECs account for the effects of the strong annihilation channels below the $N\bar{N}$ threshold~\cite{Carbonell:2023onq}. Near the $N\bar{N}$ threshold, the energy dependence from these annihilation channels is smooth and thus can be approximated by constants at LO~\cite{Dong:2020hxe}. 
The LSE for the strong-Coulomb scattering amplitude $\bm{T}_{SC}$ with the LO constant strong potential can be reduced to an algebraic equation,
\begin{align}
    \bm{T}_{SC}(E)&=\bm{V}_S(\Lambda)+\bm{V}_S(\Lambda)\bm{G}_C(E, \Lambda) \bm{T}_{SC}(E)\nonumber\\
    &=\left[\bm{I}-\bm{V}_S(\Lambda)\bm{G}_C(E, \Lambda)\right]^{-1}\bm{V}_S(\Lambda),
 \label{Eq_TSCmatrix}
\end{align}
where $\bm{G}_C(E, \Lambda)=\mathrm{diag}(G_{C11},G_{C22})$ is the Green's function regularized by a sharp cutoff $\Lambda$, with the nonvanishing diagonal matrix elements~\cite{Kong:1999sf,Konig:2015aka,Braaten:2017kci}
\begin{align}
    G_{C 11}(E, \Lambda) &= -\frac{\mu_p \Lambda}{\pi^2}-\frac{\alpha \mu_p^2}{\pi}\left(\ln \frac{\Lambda}{\alpha \mu_p}-\gamma_E\right)-\frac{\mu_p}{2 \pi} \kappa_p(E), \label{eq:GC11}\\
    G_{C 22}(E, \Lambda) &= -\frac{\mu_n \Lambda}{\pi^2}-i \frac{\mu_n}{2 \pi} k_n(E),
\end{align}
where $\mu_n=m_n/2$ is the reduced mass of $n$ and $\bar n$, $\gamma_E$ is the Euler constant, $k_n=\sqrt{2\mu_n(E-2m_n)}$ is the c.m. momentum of the neutron, and $\kappa_p = 2\alpha\mu_p\left[\mathrm{ln}(ix)+{1}/{(2ix)}-\psi(-ix)\right]$ with $\psi(x)$ being the digamma function. 

The $\Lambda$ dependence of $\bm{G}_C$ can be absorbed into the potential $\bm{V}_S$ through the renormalization procedure to achieve a cutoff-independent strong-Coulomb amplitude $\bm{T}_{SC}$, and the $\delta_{\mathrm{em}}$ term in $V_{S11}$ in Eq.~\eqref{Eq_VSCN} is utilized as a counterterm to absorb the logarithmic $\Lambda$ term in Eq.~\eqref{eq:GC11} from electromagnetic corrections and the isospin breaking due to the difference between $m_p$ and $m_n$. After renormalization, the $\bm{T}_{SC}$ matrix can be expressed in terms of the scattering length parameters $a_{ij}, i,j=1,2$ with a Coulomb modification as~\cite{Kong:1999sf}
\begin{align}
    \bm{T}_{SC}^{-1} = \bm{V}_S^{-1} - \bm{G}_C 
    = \frac{1}{2\pi} \bm{\mu}^{1/2}
    \begin{pmatrix}
        -\frac{1}{a_{11}} & \frac{1}{a_{12}} \\
         \frac{1}{a_{12}} & -\frac{1}{a_{22}}
    \end{pmatrix}
    \bm{\mu}^{1/2} - \bm{G}_C^R,
    \label{Eq_TSCsl}
\end{align}
where $\bm{\mu} =$ diag($\mu_p, \mu_n$), and 
\begin{align}
    \bm{G}_C^R = \begin{pmatrix}
        \frac{-\mu_p}{2\pi} \kappa_p(E) & 0 \\
        0 & \frac{-i \mu_n}{2\pi} k_n(E)
    \end{pmatrix}
\end{align}
is the renormalized Green's function. The physical scattering length of the $i$ th channel, $a_{ii, \, \text{eff}}$, which contains both Coulomb modification and inelasticity, is defined by the strong-Coulomb scattering amplitude at the threshold as~\cite{Sakai:2020psu}
\begin{align}
a_{ij, \, \text{eff}} =- 
 \frac{1}{2\pi} (\bm{\mu}^{1/2} \bm{T}_{SC}(E=2m_i) \bm{\mu}^{1/2})_{ij},
\label{Eq_aNN_eff}
\end{align}
where $2m_i$ is the threshold of channel $i$.

\begin{figure}[tb]
    \subfigure[]{
    \includegraphics[scale=0.6]{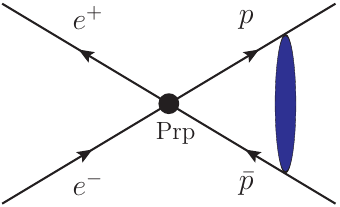}\label{diag_a}}
    \subfigure[]{
    \includegraphics[scale=0.6]{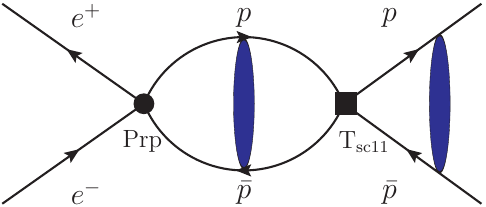}\label{diag_b}}
    \subfigure[]{
    \includegraphics[scale=0.6]{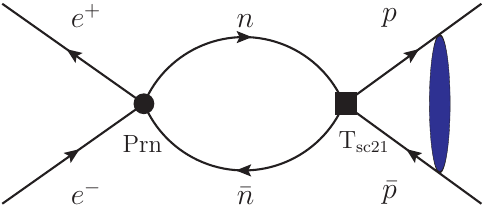}\label{diag_c}}\\
    \subfigure[]{
    \includegraphics[scale=0.6]{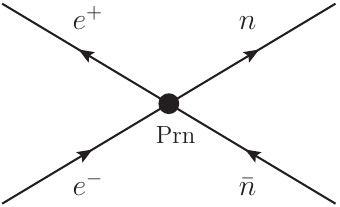}\label{diag_d}}
    \subfigure[]{
    \includegraphics[scale=0.6]{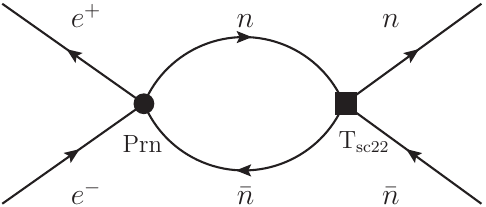}\label{diag_e}}
    \subfigure[]{
    \includegraphics[scale=0.6]{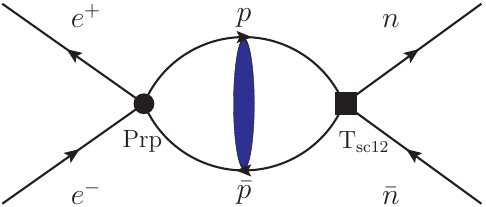}\label{diag_f}}
    \caption{Diagrams for the $e^+e^- \to p\bar{p}$ (a-c) and $e^+e^- \to n\bar{n}$ (d-f). The filled circles stand for the pointlike production sources $\mathrm{Prp}$ and $\mathrm{Prn}$ for the $p\bar{p}$ and $n\bar{n}$ production. The filled squares stand for the $p\bar{p}$ and $n\bar{n}$ interactions described by the solution of the LSE in Eq.~\eqref{Eq_TSCmatrix}. The blue blob represents infinite Coulomb-photon exchanges between proton and antiproton shown in Fig.~\ref{fig:Coulomb_resum}. The Coulomb-photon resummation between the final-state proton and antiproton is taken into account by the $W_C$ factor in the amplitude for the $e^+e^- \to p\bar{p}$ as given in Eq.~\eqref{Eq_App}. The amplitude for the $e^+e^- \to n\bar{n}$ is given in Eq.~\eqref{Eq_Ann}.}
    \label{Figs_Feynmandiags}
\end{figure}

The diagrams for the $e^+e^-\to p\bar{p}$ and $e^+e^-\to n\bar{n}$ processes are shown in Fig.~\ref{Figs_Feynmandiags}. 
Since the $N\bar N$ pairs can be produced in $S$ wave, in the near-threshold region, the $N\bar{N}$ production vertices can be approximated by cutoff-dependent constants at LO, i.e., $\mathrm{Prp}(\Lambda)$ and $\mathrm{Prn}(\Lambda)$, for the $p\bar{p}$ and $n\bar{n}$ productions, respectively. Consequently, 
the near-threshold production amplitudes for the $e^+e^- \to p\bar{p}$ and $e^+e^- \to n\bar{n}$ processes can be written as
\begin{align}
    \mathcal{M}[e^+e^- \to p\bar{p}]  &= \mathrm{Prp}(\Lambda)\times[1 +  G_{C11}(\Lambda)\times T_{SC 11}]\times W_C + \mathrm{Prn}(\Lambda)\times G_{C22}(\Lambda)\times T_{SC 21}\times W_C\nonumber\\
     &= \mathrm{Prp}(\Lambda)\times[1 +  G_{C11}(\Lambda)\times T_{SC 11} + R_{n/p}\times G_{C22}(\Lambda)\times T_{SC 21}]\times W_C,\label{Eq_App} \\
    \mathcal{M}[e^+e^- \to n\bar{n}] &= \mathrm{Prn}(\Lambda)\times [1 + G_{C22}(\Lambda)\times T_{SC 22}] +\mathrm{Prp}(\Lambda)\times G_{C11}(\Lambda)\times T_{SC 12}\nonumber\\ &= \mathrm{Prp}(\Lambda)\times [R_{n/p} + R_{n/p}\times G_{C22}(\Lambda)\times T_{SC 22} + G_{C11}(\Lambda)\times T_{SC 12}],\label{Eq_Ann}
\end{align}
where $T_{SC ij}$ are the $\bm{T}_{SC}$ matrix elements, and $R_{n/p}\equiv \mathrm{Prn}/\mathrm{Prp}$. The
three first terms of Eqs.~\eqref{Eq_App} and \eqref{Eq_Ann} are represented by the
diagrams (a), (b), (c) and (d), (e), (f) of Fig.~\ref{Figs_Feynmandiags}, respectively. The parameters Prp and Prn should be cutoff dependent, $\sim 1/\Lambda$, to absorb the $\Lambda$ dependence in the intermediate Green's functions through the multiplicative renormalization~\cite{Braaten:2005jj}. 
Since we will use the full form of the above two amplitudes, subleading cutoff dependence exists from terms with solely  Prp$(\Lambda)$ and Prn$(\Lambda)$ without Green's functions. 
We will take the cutoff $\Lambda$ to be in the range between 2.0 and 2.6~GeV, large enough to serve as a hard scale, in the following analysis.

The BESIII~\cite{BESIII:2021rqk} and SND~\cite{Achasov:2024pbk} data for cross sections are fitted by
\begin{align}
    \sigma[e^+e^- \to p\bar{p}] & = \frac{\vert \vec{p}_p \vert }{16 \pi E^2 |\vec{p}_e|}  \vert \mathcal{M}[e^+e^- \to p\bar{p}]\vert^2,
    \label{Eq_sigma_pp}\\
    \sigma[e^+e^- \to n\bar{n}] & = \frac{\vert \vec{p}_n \vert }{16 \pi E^2 |\vec{p}_e|} \vert \mathcal{M}[e^+e^- \to n\bar{n}] \vert^2,
    \label{Eq_sigma_nn}
\end{align}
where $|\vec{p}_p|$ and $|\vec{p}_n|$ are the c.m. momenta of the final-state proton and neutron, respectively, and $|\vec{p}_e|$ is the c.m. momentum of the initial-state electron. There are seven free parameters in the fit: $a_1$, $a_2$, $b_1$, $b_2$, $c_1$, $\mathrm{Prp}^2$, and $R_{n/p}$. The parameter space is constrained by unitarity, which requires the imaginary parts of the diagonal elements of the $\bm{T}$ and $\bm{V}_S$ matrices to be negative, i.e., Im$T_{ii} < 0$ and $a_2 < 0$~\cite{Badalian:1981xj,Sakai:2020psu}.

\section{Results from fits to BESIII and SND data}\label{Sec:Results}

\begin{figure}[tb]
% \subfigure[]
% {\includegraphics[width=0.496\textwidth]{figs/sigma_L2_c1_5p.pdf}}
\subfigure[]
{\includegraphics[width=0.496\textwidth]{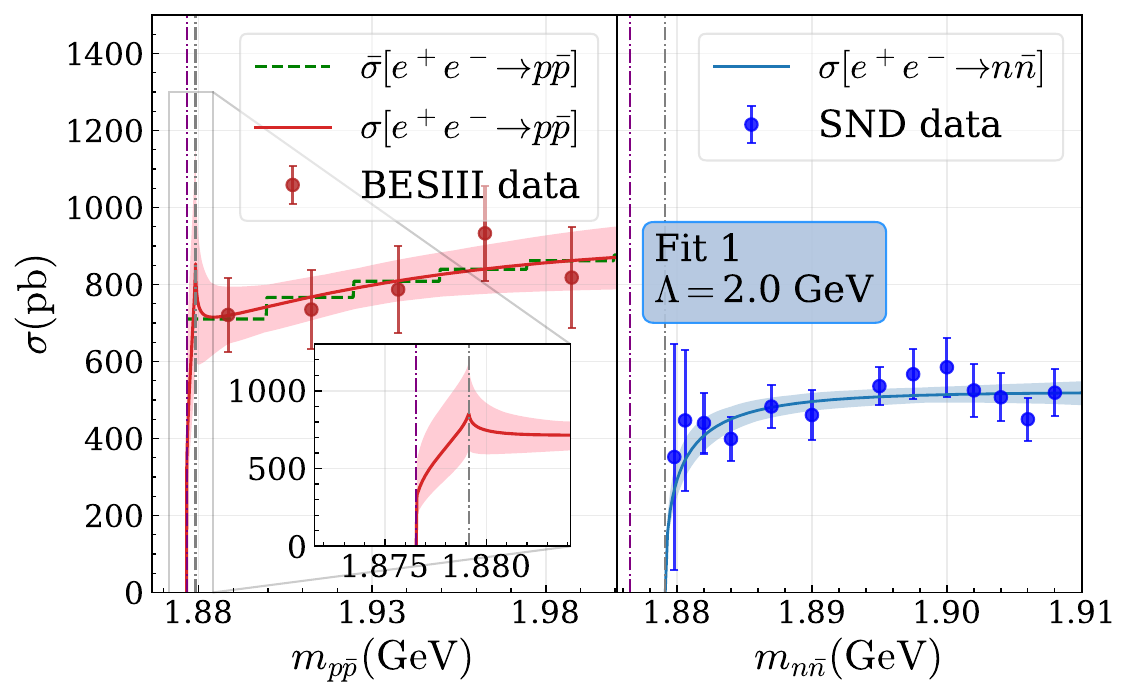}}
\subfigure[]
{\includegraphics[width=0.496\textwidth]{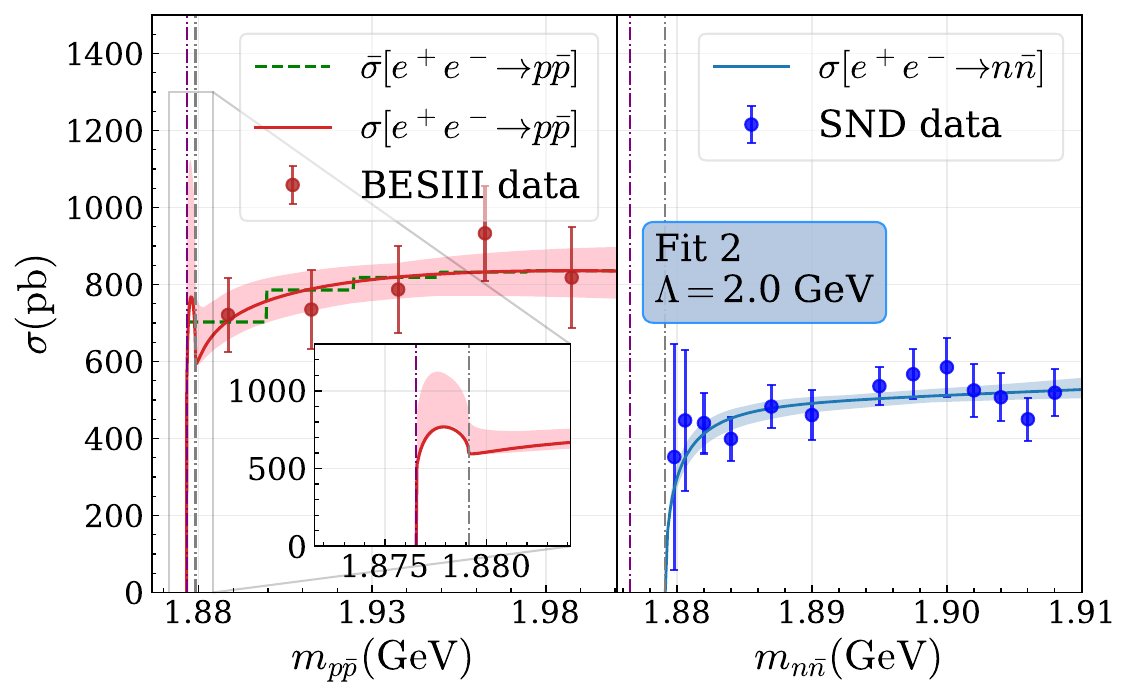}}
\caption{Comparison between the best fitted cross sections in (a) Fit 1 and (b) Fit 2 for $e^+e^- \to p\bar{p}$ and $e^+e^- \to n\bar{n}$ with $\Lambda=2.0$~GeV and the experimental data. The histograms (green dashed line) show the best fits for $e^+e^-\to p\bar{p}$ with the cross sections averaged in each bin, and the red and blue solid curves are the corresponding continuous $e^+e^- \to p\bar{p}$ and $e^+e^- \to n\bar{n}$ cross sections, respectively. The bands are the 1$\sigma$ error regions. The insets enlarge the $e^+e^- \to p\bar{p}$ line shapes in the very near-threshold region. The vertical dot-dashed lines denote the $p\bar{p}$ (purple) and $n\bar{n}$ (gray) thresholds. The data points with error bars are taken from Table 4 in Ref.~\cite{BESIII:2021rqk} for $e^+e^- \to p\bar{p}$ and Table I in Ref.~\cite{Achasov:2024pbk} for $e^+e^- \to n\bar{n}$. The purple and gray vertical dot-dashed lines denote the thresholds of the first and second channels, respectively.}
\label{Fig_sigmaFit_L2}
\end{figure}

In this section, we give the fitted near-threshold line shapes for the $N\bar{N}$ production cross sections and the poles of the $N\bar{N}$ scattering amplitudes. As the $p\bar{p}$ near-threshold production cross section data~\cite{BESIII:2021rqk} are given in bins of the $p\bar{p}$ invariant mass and the width of each bin is about 25~MeV, the cross section in Eq.~\eqref{Eq_sigma_pp} is averaged for each bin to take the binning into account in our fit. 
The averaged $p\bar{p}$ and $n\bar{n}$ production cross sections in Eqs.~\eqref{Eq_sigma_pp} and \eqref{Eq_sigma_nn} are utilized to simultaneously fit to the 5 datum points below 2.0~GeV given by BESIII~\cite{BESIII:2021rqk} and the 13 data points below 1.91~GeV measured by the SND collaboration~\cite{Achasov:2024pbk}, respectively. 
The momenta of the proton and neutron are smaller than $0.3~\rm{GeV}$, thus the NREFT treatment is well justified. We randomly generate $3 \times 10^4$ sets of initial values of the parameters constrained by unitarity to fit the data, and minimize the $\chi^2$ function using the MINUIT algorithm~\cite{James:1975dr,iminuit,iminuit.jl}. 
 
We find two fits with similar quality, denoted as Fit 1 ($\chi^2/{\rm d.o.f.} = 0.54$) and Fit 2 ($\chi^2/{\rm d.o.f.} = 0.58$).
They have the distinct features such that Fit 1 produces a sharp cusp at the $n\bar{n}$ threshold in the $p\bar{p}$ cross section and Fit 2 produces a bump between the $p\bar p$ and $n\bar n$ thresholds.
The $n\bar{n}$ cross section line shapes are almost the same in the two fits---there is no any prominent near-threshold peak but the cross section rises immediately above threshold to give an almost flat distribution, which differs from the phase space distribution drastically. The best fitted line shapes for $\Lambda$ varying between 2.0 and 2.6~GeV are almost the same. 
A comparison of the best fits with the experimental data for $\Lambda=2.0$~GeV is shown in Fig.~\ref{Fig_sigmaFit_L2}. The near-threshold data are well described by our coupled-channel formalism.
% The fitted line shapes with different $\Lambda$ and the corresponding fitted parameters can be found in Appendix~\ref{Appen:results_Lambda_compare}. The Coulomb-modified  scattering lengths in Eq.~\eqref{Eq_TSCsl} obtained with $\Lambda=2.0$~GeV extracted from the fits are listed in Table~\ref{Tab_sl}, and the $\Lambda$ dependence are shown in Fig.~\ref{Fig_slvsL}. The changes in the scattering lengths as $\Lambda$ varying from 2.0 to 2.6 GeV is about 20\%, which is indeed subleading as expected. The difference between the values of $a_{11}$ and $a_{22}$ is due to isospin breaking effects.

\begin{figure}[tb]
\subfigure[]
{\includegraphics[width=0.496\textwidth]{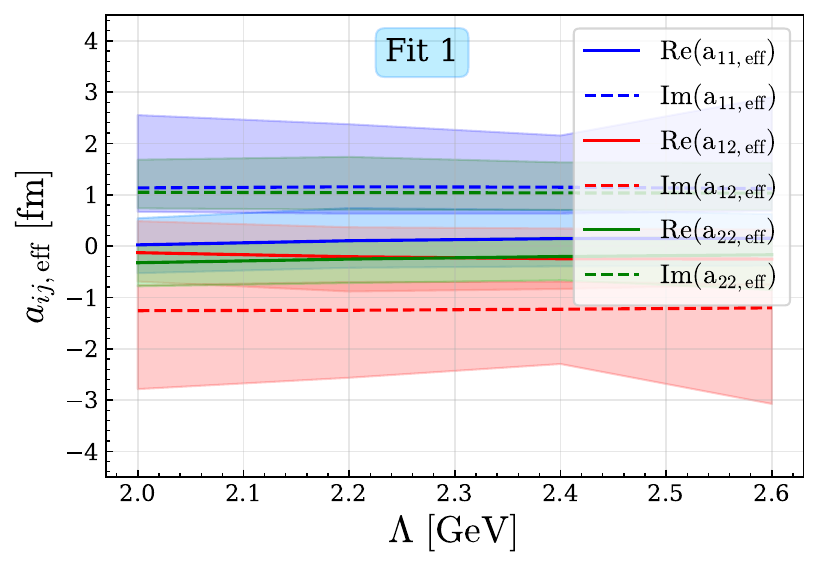}}
\subfigure[]
{\includegraphics[width=0.496\textwidth]{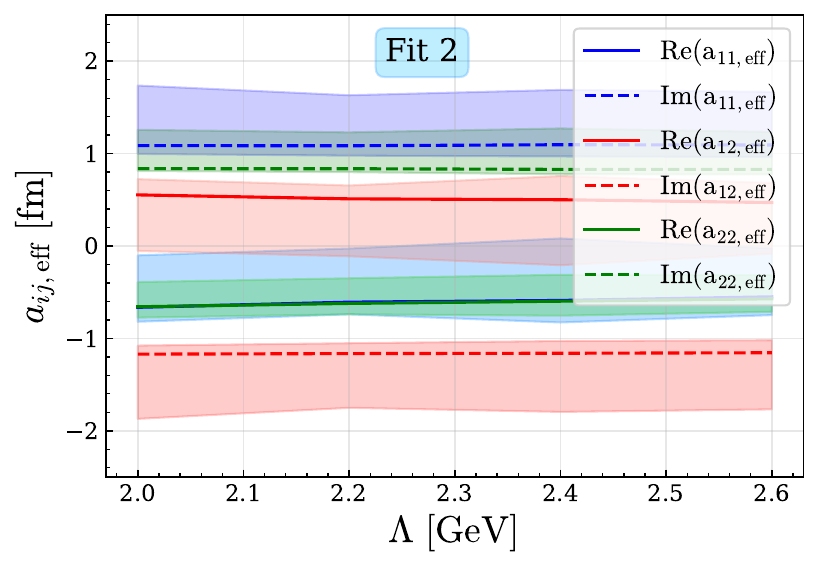}}

\caption{Effective scattering length as a function of $\Lambda$ extracted from (a) Fit 1 and (b) Fit 2. The bands represent the $1\sigma$ errors propagated from the errors of the data.}
\label{Fig_slvsL}
\end{figure}

The Coulomb-modified effective scattering lengths in Eq.~\eqref{Eq_aNN_eff} for $\Lambda=2.0-2.6$~GeV extracted from the fits are listed in Table~\ref{Tab_sl}. The central values of the effective scattering lengths are the averaged values of the central values for different $\Lambda$, and the uncertainties are evaluated as the difference between the central values and the extreme values of the effective scattering lengths for all different $\Lambda$ within the 1$\sigma$ errors propagated from the data. The uncertainties include both the statistical errors from the data and the systematic errors from the variation of $\Lambda$. \footnote{Here the uncertainties are evaluated in a conservative way. The statistical and systematic uncertainties can also be evaluated separately by taking the central values and statistical errors at a certain $\Lambda$, e.g., $\Lambda=2.2$~GeV, and taking the variation of the central values for different $\Lambda$ as the systematic errors. Since the uncertainties are completely dominated by the statistical ones, both ways give similar results.} As can be seen from Fig.~\ref{Fig_slvsL}, the variations in the effective scattering lengths as $\Lambda$ varying from 2.0 to 2.6 GeV are marginal, which is expected as the residual cutoff dependence should be subleading effects. The difference between the values of $a_{11,\,\text{eff}}$ and $a_{22,\,\text{eff}}$ is due to isospin breaking effects. The value of the effective scattering length $a_{11,\,\text{eff}}$ is comparable with the $p\bar{p}$ Coulomb-modified scattering length in the $\leftindex^3 S_1$--$\leftindex^3 D_1$ channel,  $-0.933(45)+i 0.604(51)~\rm{fm}$ extracted from the energy level shift of the antiprotonic hydrogen~\cite{Gotta:1999vj,Gotta:2004rq,Carbonell:2023onq} with the Trueman formula~\cite{Trueman:1961zza}.\footnote{Here we use an opposite sign convention for the scattering length compared to that in Refs.~\cite{Gotta:2004rq,Carbonell:2023onq}.}

% \begin{table}[tb]
% \caption{\label{Tab_sl} Scattering lengths obtained with $\Lambda=2.0$~GeV. The uncertainties are the $1\sigma$ errors propagated from the statistical errors of the data.}

% \begin{ruledtabular}
% \begin{tabular}
% {l|ccc}

%         Fit
%         & $a_{11}~\rm{[fm]}$
%         & $a_{12}~\rm{[fm]}$
%         & $a_{22}~\rm{[fm]}$
%         \\[3pt]     
% \hline

%   1 &$-0.25_{-0.01}^{+0.01}+i (0.03_{-0.01}^{+0.01})$ &$0.28_{-0.01}^{+0.01}-i( 0.00_{-0.01}^{+0.01})$ &$-0.29_{-0.02}^{+0.01}+i (0.03_{-0.01}^{+0.01})$ \\[3pt]
 
%    2 &$-0.29_{-0.01}^{+0.00}+i (0.02_{-0.00}^{+0.00})$ &$0.33_{-0.01}^{+0.00}+i (0.00_{-0.00}^{+0.00})$ &$-0.31_{-0.01}^{+0.00}+i(0.03_{-0.00}^{+0.01})$ \\[3pt] 
% \end{tabular}
% \end{ruledtabular}
% \end{table}

\begin{table}[tb]
\caption{\label{Tab_sl} The Coulomb-modified effective scattering lengths for the $N\bar{N}$ coupled-channel scattering with the cutoff $\Lambda =2.0-2.6~\rm{GeV}$. The central values are the average values of those corresponding to different $\Lambda$. The uncertainties are the $1\sigma$ errors propagated from the statistical uncertainties of the data and the systematic error from the variation of the cutoff.}

\begin{ruledtabular}
\begin{tabular}
{l|ccc}

        Fit
        & $a_{11,\,\text{eff}}~\rm{[fm]}$
        & $a_{12,\,\text{eff}}~\rm{[fm]}$
        & $a_{22,\,\text{eff}}~\rm{[fm]}$
        \\[3pt]     
\hline

  1 &$0.10_{-0.63}^{+0.63}+i (1.14_{-0.51}^{+1.76})$ &$-0.21_{-0.68}^{+0.69}-i(1.24_{-0.55}^{+1.84})$ &$-0.24_{-0.62}^{+0.41}+i (1.04_{-0.34}^{+0.70})$ \\[3pt]
 
   2 &$-0.60_{-0.23}^{+0.68}+i (1.09_{-0.13}^{+0.65})$ &$0.51_{-0.72}^{+0.25}-i (1.16_{-0.14}^{+0.71})$ &$-0.61_{-0.16}^{+0.30}+i(0.83_{-0.06}^{+0.44})$ \\[3pt] 
\end{tabular}
\end{ruledtabular}
\end{table}

% \begin{table}[tb]
% \caption{\label{Tab_sl} \blue{The Coulomb-modified effective scattering lengths for the $N\bar{N}$ coupled-channel scattering with the cutoff $\Lambda =2.0-2.6~\rm{GeV}$. The central values are the average values of those corresponding to different $\Lambda$. The uncertainties are the $1\sigma$ errors propagated from the statistical uncertainties of the data and the systematic error from the variation of the cutoff.}}

% \begin{ruledtabular}
% \begin{tabular}
% {l|ccc}

%         Fit
%         & $a_{11,\,\text{eff}}~\rm{[fm]}$
%         & $a_{12,\,\text{eff}}~\rm{[fm]}$
%         & $a_{22,\,\text{eff}}~\rm{[fm]}$
%         \\[3pt]     
% \hline

%   1 &$0.10_{-0.53-0.08}^{+0.64+0.05}+i (1.15_{-0.52-0.03}^{+1.22+0.00})$ 
  
%   &$-0.21_{-0.68-0.04}^{+0.57+0.08}-i(1.25_{-0.55-0.01}^{+1.32+0.05})$ 
  
%   &$-0.25_{-0.47-0.08}^{+0.35+0.08}+i (1.04_{-0.33-0.01}^{+0.70+0.01})$ \\[3pt]
 
%    2 &$-0.61_{-0.13-0.05}^{+0.58+0.06}+i (1.08_{-0.11-0.00}^{+0.55+0.02})$ &$0.51_{-0.62-0.04}^{+0.14+0.04}-i (1.16_{-0.11-0.01}^{+0.59+0.00})$ &$-0.62_{-0.12-0.04}^{+0.27+0.05}+i(0.84_{-0.04-0.01}^{+0.40+0.00})$ \\[3pt] 
% \end{tabular}
% \end{ruledtabular}
% \end{table}

\begin{figure}[tb]
% \subfigure[]
% {\includegraphics[width=0.496\textwidth]{figs/poles_2c_fit1_L2_5p.pdf}}
\subfigure[]
{\includegraphics[width=0.496\textwidth]{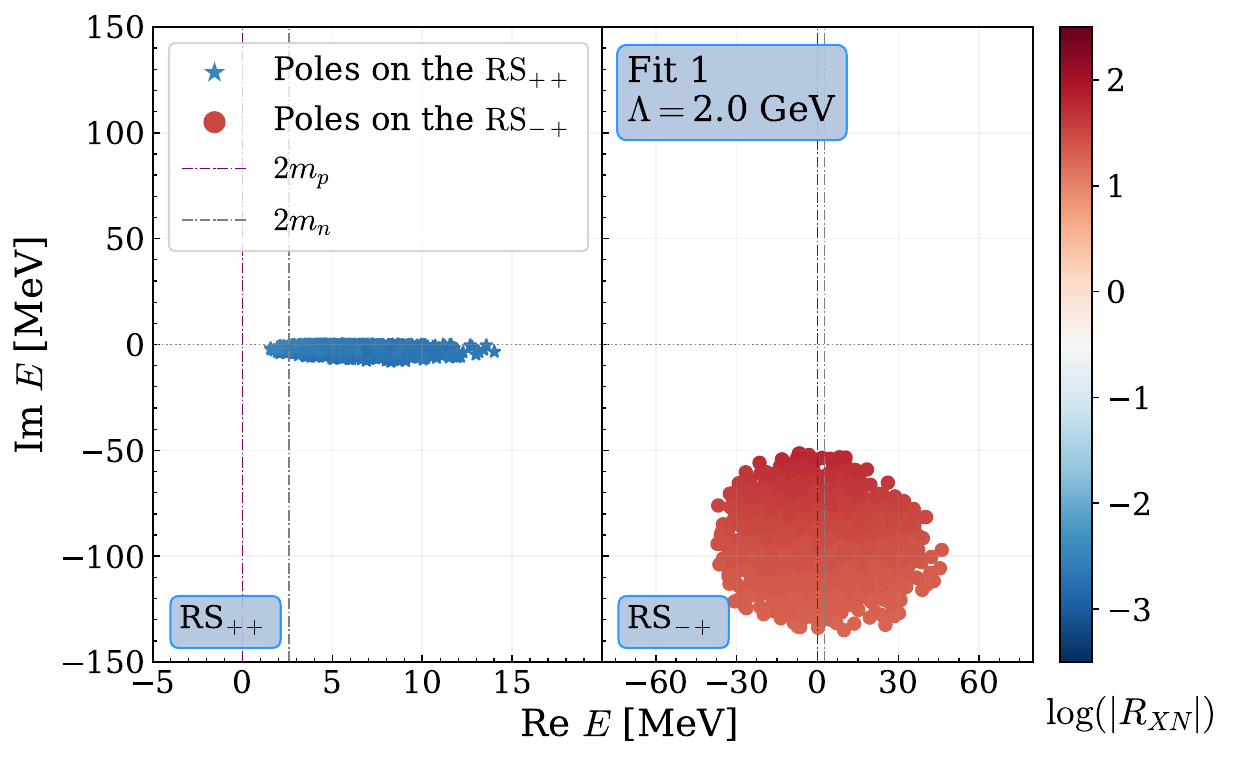}}
\subfigure[]
{\includegraphics[width=0.496\textwidth]{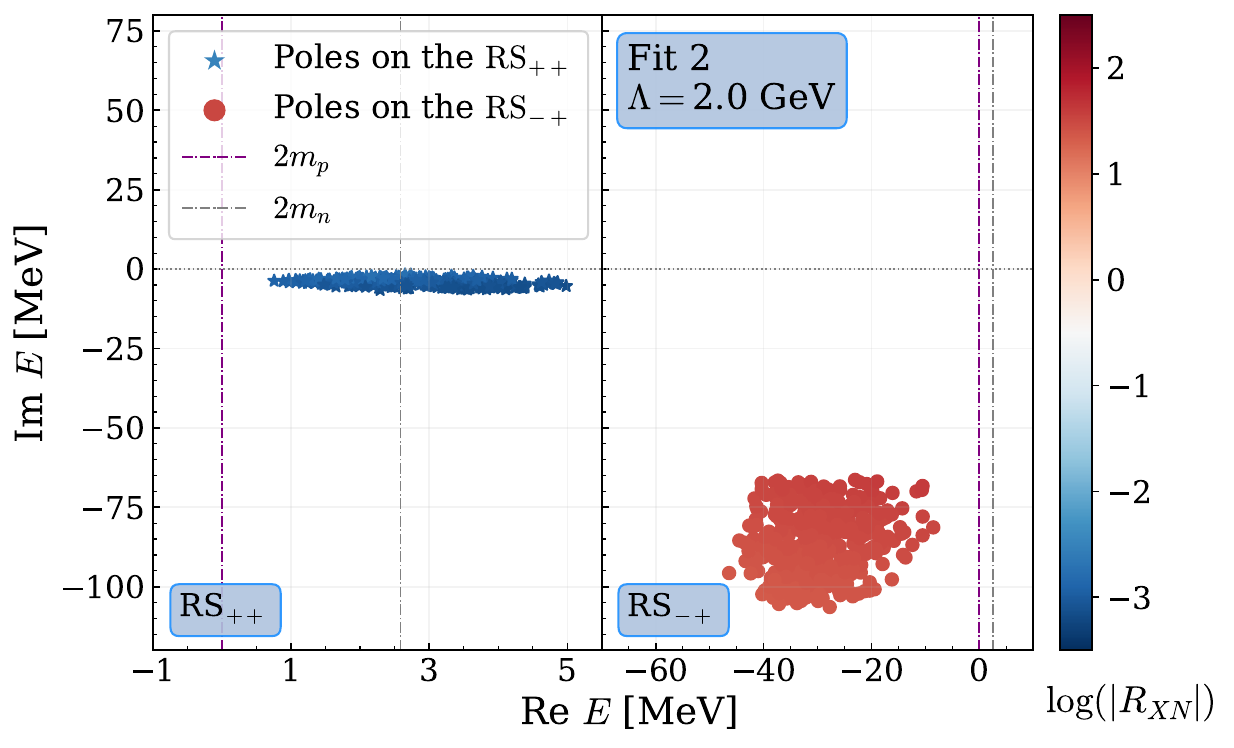}}
\caption{Pole positions on the first and second RSs from (a) Fit 1 and (b) Fit 2 using parameters in the 1$\sigma$ error range for $\Lambda=2.0$~GeV. The logarithm of the absolute value of ratio between the isovector and isoscalar residues of each pole, $R_{XN}=g_{X,I=1}/g_{X,I=0}$, is encoded in color.}
\label{Fig_poleFit_L2}
\end{figure}

\begin{table}[tb]
\caption{\label{Tab_pole}Pole positions with different $\Lambda$. $E_1$ and $E_2$ represent the pole positions on the RS$_{++}$ and RS$_{-+}$, respectively. The uncertainties are the $1\sigma$ errors propagated from the statistical uncertainties of the data.}

\begin{ruledtabular}
\begin{tabular}
{l|c cc}
        $\Lambda~\rm{[GeV]}$
        & Fit
        & $E_1~[\mathrm{MeV}] (\mathrm{RS_{++}})$ 
        & $E_2~[\mathrm{MeV}] (\mathrm{RS_{-+}})$
        \\[3pt]     
\hline
  \multirow{2}{*}{2.0} &1 &$1882.1_{-4.2}^{+9.5}-i (2.0_{-2.0}^{+5.5})$ &$1877.3_{-40.1}^{+47.5}-i( 89.3_{-44.3}^{+53.8})$  \\[3pt]

  &2&$1877.7_{-0.5}^{+3.9}-i(5.8_{-4.1}^{+0.9})$ &$1826.3_{-5.2}^{+40.6}-i(84.1_{-18.9}^{+21.7})$\\[3pt]
  \hline
  \multirow{2}{*}{2.2} &1 &$1882.1_{-4.4}^{+7.6}-i(1.2_{-1.2}^{+5.4})$ &$1879.3_{-42.6}^{+64.2}-i(100.0_{-58.7}^{+57.9})$  \\[3pt]

  &2&$1878.1_{-0.9}^{+4.1}-i(5.9_{-4.7}^{+1.2})$ &$1822.1_{-1.2}^{+48.7}-i(93.6_{-20.7}^{+26.2})$ \\[3pt]
    \hline
  \multirow{2}{*}{2.4} &1 &$1882.2_{-4.5}^{+9.2}-i(0.8_{-0.8}^{+5.0})$ &$1883.0_{-47.3}^{+71.6}-i(111.5_{-56.2}^{+67.5})$  \\[3pt]

  &2&$1878.3_{-1.1}^{+3.8}-i(5.9_{-4.3}^{+1.2})$ &$1813.3_{-0.0}^{+52.1}-i(105.0_{-23.9}^{+27.3})$ \\[3pt]
    \hline
  \multirow{2}{*}{2.6} &1 &$1882.4_{-3.8}^{+7.7}-i(0.6_{-0.6}^{+4.9})$ &$1889.2_{-51.9}^{+72.4}-i(124.7_{-59.9}^{+65.2})$   \\[3pt]

  &2&$1878.7_{-1.5}^{+4.0}-i(6.0_{-4.6}^{+1.4})$ &$1809.2_{-4.6}^{+57.3}-i(114.8_{-27.6}^{+31.3})$\\[3pt]
\end{tabular}
\end{ruledtabular}
\end{table}

% Although the magnitudes of the scattering lengths $a_{11}$ and $a_{22}$ are small, the large $1/|a_{12}|\gg \sqrt{2m_p(m_n-m_p)}$ indicates strong channel coupling and thus the possible near-threshold $N\bar{N}$ $\leftindex^3 {S}_1$ 
% resonant states~\cite{Zhang:2024qkg}. These resonant states appear as poles on the RSs of the complex energy $E$ plane for the $\bm{T}_{SC}(E)$ matrix in Eq.~\eqref{Eq_TSCmatrix}. 

We find two near-threshold poles for both Fit 1 and Fit 2. 
One pole is located on the first (physical) RS ($\mathrm{RS}_{++}$), and the other one is on the second RS ($\mathrm{RS}_{-+}$), where $\mathrm{RS}_{\pm \pm}$ is quoted to indicate the RSs where the poles are located, with the first and second signs in the subscript representing the signs of Im$k_p$ and Im$k_n$, respectively. 
The pole on the RS$_{++}$ also has a shadow pole~\cite{Eden:1964zz} on RS$_{+-}$, which is far from the physical region and will not be discussed in the following. 
The uncertainty of the pole positions for each fit is computed using $3\times 10^4$ parameter sets within the $1\sigma$ error region for $\Lambda=2.0 - 2.6$~GeV, and the results for $\Lambda=2.0$ GeV are shown in Fig.~\ref{Fig_poleFit_L2}. 

The pole positions for different $\Lambda$ values are similar, as listed in Table~\ref{Tab_pole}, where $E_1$ and $E_2$ denote the positions of the poles on the first and second RSs, respectively. For comparison, the pole positions with different $\Lambda$ for Fit 1 and Fit 2 are given in Figs.~\ref{Fig_poleFit1} and \ref{Fig_poleFit2}, respectively, in the Appendix~\ref{Appen:results_Lambda_compare}.
Taking the uncertainties of the pole positions to cover the variations with $\Lambda\in [2.0,2.6]$~GeV and the mean values of the central values listed in Table~\ref{Tab_pole} as the central values, we obtain for Fit 1
\begin{align}
    E_1 = 1882.2_{-4.5}^{+9.4} - i (1.2_{-1.2}^{+6.3})~{\rm MeV},\qquad
    E_2 = 1882.2_{-46.5}^{+79.4} - i (106.4_{-65.1}^{+83.5})~{\rm MeV},
\end{align}
and for Fit 2
\begin{align}
    E_1 = 1878.2_{-1.0}^{+4.5} - i (5.9_{-4.7}^{+1.5})~{\rm MeV},\qquad
    E_2 = 1817.7_{-13.1}^{+53.1} - i (99.4_{-34.2}^{+46.7})~{\rm MeV}.
\end{align}

The pole on RS$_{++}$ in both fits is rather close to the $p\bar p$ threshold at 1876.27~MeV and $n\bar n$ threshold at 1879.13~MeV, with the one in Fit 1 above the $n\bar n$ threshold and the one in Fit 2 between the two thresholds.
Such difference in pole locations causes the different behaviors of the $p\bar p$ cross section near the $n\bar n$ threshold in the two fits~\cite{Zhang:2024qkg}.
The pole would become a bound state pole below the $p\bar{p}$ threshold without an imaginary part if the lower annihilation channels are turned off~\cite{Dmitriev:2015qyt,Xiao:2024jmu}.

The pole on RS$_{-+}$ is much deeper in the complex plane compared with the one on RS$_{++}$, and has smaller impact on the line shapes, thus bearing larger uncertainties from fitting to line shape data.
One also notices that the pole location differs significantly in the two fits, with the one in Fit 2 farther away from the thresholds than that in Fit 1. 
 
Since the isospin of the $N\bar{N}$ system can be either 0 or 1, the existence of two distinct poles at LO of NREFT indicates the presence of one isoscalar and one isovector $N\bar N$ molecular states. This expectation is confirmed by the couplings of the poles to the isoscalar and isovector channels.
The effective couplings of the two near-threshold resonant states to $N\bar{N}$ can be derived from the residues of the $\bm{T}$ matrix at the corresponding pole positions as
 \begin{align}
     g_{Xi}g_{Xj}=\lim\limits_{E\to E_X}(E-E_X)T_{ij}(E)
 \end{align}
where $X=1,2$ correspond to the two poles on the first and second RSs, respectively, and $i,j=1,2$ are the channel indices. 
The couplings to the $I=0,1$ channels for the state corresponding to the pole $X$ are
\begin{align}
    g_{X,I=0}=-\frac{1}{\sqrt{2}}(g_{X1}+g_{X2}),\qquad 
    g_{X,I=1}=-\frac{1}{\sqrt{2}}(g_{X1}-g_{X2}).
\end{align}
The magnitude of the ratio $R_{XN}\equiv g_{X,I=1}/g_{X,I=0}$ measures the relative strength between couplings for the $X$ state to the isovector and isoscalar channels. 
The $\mathrm{log}(|R_{XN}|)$ value is encoded in colors in Fig.~\ref{Fig_poleFit_L2} (see also Figs.~\ref{Fig_poleFit1} and \ref{Fig_poleFit2} in the Appendix~\ref{Appen:results_Lambda_compare}). 

For the pole on RS$_{++}$, the absolute value of its coupling to the isoscalar channel is about 2 to 3 orders of magnitude larger than that to the isovector channel. Therefore, the state is predominantly an isoscalar state, which is consistent with the previous studies~\cite{Kang:2013uia,Haidenbauer:2014kja,Dmitriev:2015qyt,Haidenbauer:2015yka,Milstein:2018orb,Xiao:2024jmu}. 

The pole on RS$_{-+}$ couples stronger to the isovector channel, as its coupling to the isovector channel is about 1 to 2 orders of magnitude larger than that to the isoscalar channel. Previous studies also did not find any bound states in the isovector channel~\cite{Kang:2013uia,Haidenbauer:2014kja,Dmitriev:2015qyt,Haidenbauer:2015yka,Milstein:2018orb,Xiao:2024jmu}. Our result that the isovector pole is on RS$_{-+}$ instead of RS$_{++}$ is consistent with these studies. 

\section{Results with constraint from antiprotonic hydrogen}\label{results with fixed a11} 

To further check the consistency of our results with the experimental results of the antiprotonic hydrogen~\cite{Gotta:1999vj,Gotta:2004rq,Carbonell:2023onq}, we perform another fit with the Coulomb-modified $p\bar p$ scattering length fixed to $a_{11,\rm{eff}}=-0.933(45)+i 0.604(51)~\rm{fm}$ obtained from the atomic measurement via the Trueman relation~\cite{Trueman:1961zza}. 
As given in Eq.~\eqref{Eq_aNN_eff}, the strong-Coulomb scattering amplitude for the c.m. energy at the $p\bar{p}$ threshold can be expressed in terms of the effective scattering lengths as 
\begin{align}
    \bm{T}_{SC}^{\rm{thr}} = -2 \pi\bm{\mu}^{-\frac{1}{2}} \bm{a}_{NN,\,\rm{eff}} \bm{\mu}^{-\frac{1}{2}},
\end{align}
where
\begin{align}
    \bm{a}_{NN,\,\rm{eff}} = \begin{pmatrix}
        a_{11,\,\rm{eff}}^{\rm{Re}}+i a_{11,\,\rm{eff}}^{\rm{Im}} & a_{12,\,\rm{eff}}^{\rm{Re}}+i a_{12,\,\rm{eff}}^{\rm{Im}} \\
        a_{12,\,\rm{eff}}^{\rm{Re}}+i a_{12,\,\rm{eff}}^{\rm{Im}} & \tilde{a}_{22,\,\rm{eff}}^{\rm{Re}}+i \tilde{a}_{22,\,\rm{eff}}^{\rm{Im}}
    \end{pmatrix},
    \label{Eq_aNNeff}
\end{align}
with $a_{ij,\,\rm{eff}}^{\rm{Re(Im)}}$ ($i, j=1, 2$) representing the real (imaginary) part of the effective scattering length $a_{ij,\,\rm{eff}}$. 
Here $\tilde{a}_{22,\,\rm{eff}}$ is obtained at the lower threshold $E=2m_p$, and is different from the effective scattering length of channel 2 defined in Eq.~\eqref{Eq_aNN_eff}. 
The inverse of the strong-Coulomb amplitude $\bm{T}_{SC}$ can be expressed as
\begin{align}
    \bm{T}_{SC}^{-1}(E) &= (\bm{V}_S^R)^{-1} - \bm{G}_C^R(E)\nonumber\\
    &= (\bm{T}_{SC}^{\rm{thr}})^{-1} + \bm{G}_C^R(E=2m_p) - \bm{G}_C^R(E)\nonumber\\
    &= (\bm{T}_{SC}^{\rm{thr}})^{-1} - \tilde{\bm{G}}_C^{R },
    \label{Eq_TSCsl_2nd}
\end{align}
where $\tilde{\bm{G}}_C^{R} = \bm{G}_C^R(E) - \bm{G}_C^R(E=2m_p)$. 
The amplitude $\bm{T}_{SC}$ in Eq.~\eqref{Eq_TSCsl_2nd} is the one utilized in the production amplitudes in Eqs.~\eqref{Eq_App} and \eqref{Eq_Ann}. Then the averaged $p\bar{p}$ and $n\bar{n}$ production cross sections in Eqs.~\eqref{Eq_sigma_pp} and \eqref{Eq_sigma_nn} are utilized to simultaneously fit to the same datasets in Sec.~\ref{Sec:Results}.  We again randomly generate $3 \times 10^4$ sets of initial values of the parameters constrained by unitarity to fit the data, and minimize the $\chi^2$ function using the MINUIT algorithm~\cite{James:1975dr,iminuit,iminuit.jl}.

\begin{figure}[tb]
\subfigure[]
{\includegraphics[width=0.496\textwidth]{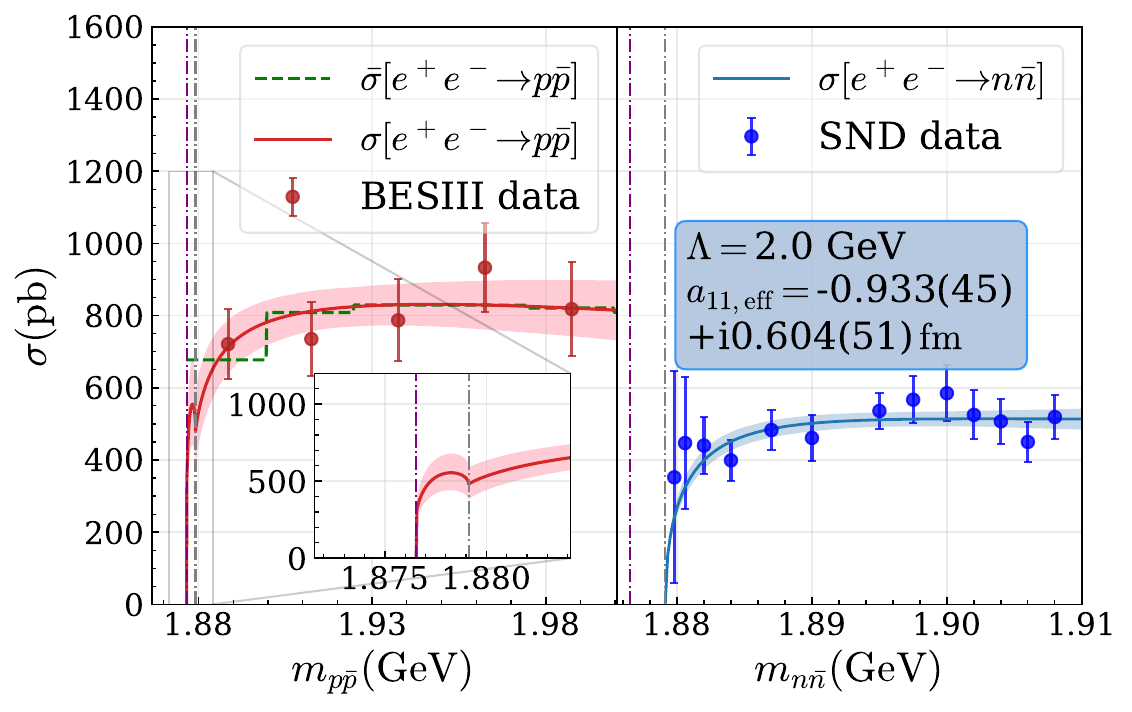}\label{Fig_sigma_fixeda11}}
\subfigure[]
{\includegraphics[width=0.496\textwidth]{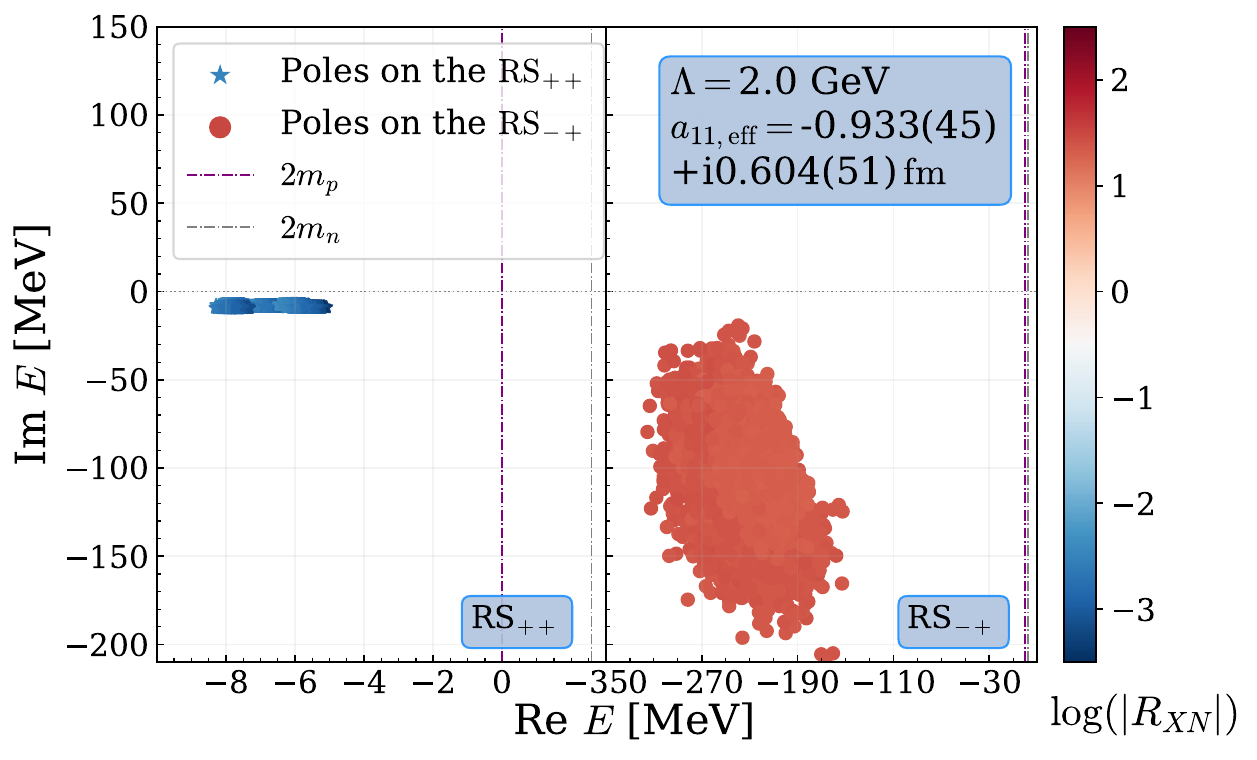}\label{Fig_poles_fixeda11}}
\caption{(a) Comparison between the best fitted cross sections for $e^+e^- \to p\bar{p}$ and $e^+e^- \to n\bar{n}$ with $\Lambda=2.0$~GeV and the experimental data; see the caption of Fig.~\ref{Fig_sigmaFit_L2}. (b) Pole positions on RS$_{++}$ and RS$_{-+}$ for $\Lambda=2.0$~GeV; see the caption of Fig.~\ref{Fig_poleFit_L2}. The errors include the $1\sigma$ errors propagated from the statistical errors of the data and those from $a_{11,\,\text{eff}}$.}
\label{Fig_sigmapoles_fixeda11}
\end{figure}

The data can be well reproduced.
The fitted line shapes of the cross sections are similar to those of Fit 2 in the previous section. A comparison of the best fit results with the experimental data and the pole positions for $\Lambda = 2.0~\rm{GeV}$ is shown in Fig.~\ref{Fig_sigmapoles_fixeda11}. There is a bump in the line shape of the $p\bar{p}$ production cross section between the two thresholds, which, however, is less prominent than that in Fit 2. The $\chi^2 /\text{d.o.f.}$, Coulomb-modified effective scattering lengths, $\mathrm{Prp}^2$, and $R_{n/p}$ with a few $\Lambda$ values are listed in Table~\ref{Tab_parametersFit_fixeda11eff}. 
It is interesting to note that the variations of the free parameters as a function of $\Lambda$ shown in Table~\ref{Tab_parametersFit_fixeda11eff} when $a_{11, \rm{eff}}$ is
fixed to that extracted from antiprotonic hydrogen are smaller than those listed in Tables~\ref{Tab_parametersFit1} and \ref{Tab_parametersFit2} where no constraint on the strong potential $V_S$ was imposed. The pole positions with different $\Lambda$ are given in Fig.~\ref{Fig_poleFit_fixeda11} in the Appendix~\ref{Appen:results_Lambda_compare}. Taking the uncertainties including those propagated from the data and $a_{11,\,\text{eff}}$, and those from the variations of the cutoff $\Lambda = 2.0 - 2.6~\rm{GeV}$, we obtain 
\begin{align}
    a_{22,\,\rm{eff}} = -0.628_{-0.084}^{+0.062} + i (0.396_{-0.090}^{+0.108})~{\rm fm}.
\end{align}
While the real part is compatible with that in Fit 2, the imaginary part is smaller. 

The pole positions on RS$_{++}$ and RS$_{-+}$ are
\begin{align}
    E_1 = 1869.7_{-1.7}^{+1.7} - i (8.5_{-1.6}^{+1.4})~{\rm MeV},\qquad
    E_2 = 1582.6_{-251.1}^{+141.6} - i (144.7_{-124.7}^{+148.2})~{\rm MeV},
\end{align}
whose central values are the mean values for different $\Lambda$ with $a_{11,\,\text{eff}}=-0.933+i 0.604$~fm.
In comparison with the Fit 2, the pole on RS$_{-+}$ are much deeper and farther from the thresholds. It is too far from the $N\bar N$ thresholds to have physical significance, and the bump is dominantly due to the pole on RS$_{++}$.
% , which could be attributed to the inaccurate near-threshold data of the $p\bar{p}$ production cross section and explains the much lower bump in the $p\bar{p}$ production cross section than that in Fit 2.}

\begin{table}[tb]
\caption{\label{Tab_parametersFit_fixeda11eff}Parameter sets from fit with $a_{11,\rm{eff}}=-0.933(45)+i 0.604(51)~\rm{fm}$~\cite{Gotta:2004rq,Carbonell:2023onq} for different $\Lambda$ values. The uncertainties include the $1\sigma$ errors propagated from the statistical uncertainties of the data and the errors propagated from those of the $a_{11,\,\text{eff}}$.}
\renewcommand{\arraystretch}{1.2}
\begin{tabular*}{\columnwidth}{@{\extracolsep\fill}lcccccccc}
\hline\hline                                  
            $\Lambda~\rm{[GeV]}$
            & $\chi^2$/d.o.f. 
            & $a_{12,\,\rm{eff}}~\rm{[fm]}$
            & $\tilde{a}_{22,\,\rm{eff}}~\rm{[fm]}$
            % & $a_{22,\,\rm{eff}}~\rm{[fm]}$
            & $\mathrm{Prp}^2$
            & $R_{n/p}$
            \\[3pt]     
\hline
      2.0 &$0.57$ 
      &$0.774_{-0.068}^{+0.089}-i(0.520_{-0.106}^{+0.080})$
      &$-0.860_{-0.107}^{+0.094}
      +i(0.475_{-0.117}^{+0.168})$ 
      % &$-0.651_{-0.060}^{+0.061}
      % +i(0.398_{-0.072}^{+0.105})$
      &$0.012_{-0.005}^{+0.020}$ &$1.020_{-0.091}^{+0.096}$\\[3pt]
      2.2 &$0.56$ 
      &$0.770_{-0.089}^{+0.063}-i(0.520_{-0.088}^{+0.089})$
      &$-0.835_{-0.103}^{+0.114}
      +i(0.471_{-0.130}^{+0.136})$ 
      &$0.013_{-0.007}^{+0.014}$ &$1.045_{-0.087}^{+0.113}$\\[3pt]
      2.4 &$0.56$ 
      &$0.768_{-0.063}^{+0.069}
      -i(0.520_{-0.071}^{+0.085})$
      &$-0.820_{-0.087}^{+0.067}
      +i(0.465_{-0.116}^{+0.089})$ 
      &$0.017_{-0.009}^{+0.015}$ &$1.071_{-0.067}^{+0.080}$\\[3pt]
      2.6 &$0.56$ 
      &$0.767_{-0.071}^{+0.081}
      -i(0.509_{-0.075}^{+0.089})$
      &$-0.794_{-0.095}^{+0.069}
      +i(0.448_{-0.133}^{+0.102})$ 
      &$0.011_{-0.005}^{+0.024}$ &$1.073_{-0.066}^{+0.103}$\\[3pt]
    \hline\hline
    \end{tabular*}
\end{table}

\section{Summary}\label{Sec:Summary}

In this work, we studied the most updated data of the near-threshold $e^+e^-\to p\bar{p}$ and $e^+e^-\to n\bar{n}$ cross sections reported by the BESIII~\cite{BESIII:2021rqk} and SND~\cite{Achasov:2024pbk} collaborations in the NREFT framework at LO. The framework includes both the $p\bar p$ and $n\bar n$ channels, with isospin breaking from the proton-neutron mass difference and the Coulomb interactions between proton and antiproton.

We found two fits with similar quality that can well describe the near-threshold experimental cross sections. 
The line shape of the $p\bar{p}$ production cross section has a sharp cusp at the $n\bar{n}$ threshold for Fit 1 and a bump between the $p\bar{p}$ and $n\bar{n}$ thresholds for Fit 2.
The Coulomb-modified scattering lengths~\cite{Kong:1999sf} for the $p\bar{p}$--$n\bar{n}$ coupled channels were obtained, and the results indicated a strong channel coupling~\cite{Zhang:2024qkg}.  
Using the LECs determined from the fits, we studied possible $N\bar{N}$ composite states in the $\leftindex^3 S_1$ partial wave, which appear as poles of the strong-Coulomb $\bm{T}_{SC}$ matrix. 
As a consequence of the strong $p\bar p$--$n\bar n$ channel coupling, two different poles exist in the orthogonal isoscalar and isovector channels. 
We obtained an isoscalar near-threshold $N\bar{N}$ quasibound state pole on the physical RS (RS$_{++}$), which lies either above the $n\bar n$ threshold (in Fit 1) or between the $p\bar p$ and $n\bar n$ thresholds (in Fit 2). 
We also found an isovector pole on the unphysical RS$_{-+}$. It is at least a few tens of MeV away from the $N\bar N$ thresholds, but its position is not well determined, differing sizably between the two fits and bearing a large uncertainty.

To further confirm the consistency between our result and the experimental result obtained from the antiprotonic hydrogen~\cite{Gotta:1999vj,Gotta:2004rq,Carbonell:2023onq}, we performed another fit with the effective $p\bar p$ scattering length fixed to that extracted from the energy level shift of the $\bar{p}H$ atom. The results are similar to the results in Fit 2. There is a bump, though less prominent than that in Fit 2, between the two thresholds in the $p\bar{p}$ production cross section. Our results update the understanding on the near-threshold structures in the $e^+e^-\to N\bar{N}$ cross sections and the possible isoscalar and isovector $N\bar{N}$ composite states with $J^{PC}=1^{--}$. 

The amplitudes we derived for the $N\bar{N}$ FSIs can also be used in analyzing near-threshold data in other processes, such as $J/\psi \to \gamma p\bar{p}$~\cite{BES:2003aic,BESIII:2010vwa}, $B^{\pm}\to K^{\pm}p\bar{p}$ decays~\cite{Belle:2002bro,Belle:2002fay,Belle:2007oni}, and so on.

\section{ACKNOWLEDGMENTS}\label{sec: ACKNOWLEDGMENTS}

We would like to thank Yan-Ping Huang who encouraged us to perform this study.
This work is supported in part by the Chinese Academy of Sciences under Grants No. XDB34030000 and No.~YSBR-101; by the National Key R\&D Program of China under Grant No. 2023YFA1606703; by the National Natural Science Foundation of China (NSFC) under Grants No.~12125507, No.~12047503, No.~12475081, No.~12075133, and No.~12361141819; by the Natural Science Foundation of Shandong province under the Grant No. ZR2022ZD26; and by Taishan Scholar Project of Shandong Province under Grant No. tsqn202103062.

\appendix
\section{FITTED RESULTS WITH \texorpdfstring{$\Lambda=2.0-2.6$} {Lambda=2.0-2.6} GeV}\label{Appen:results_Lambda_compare}

In this Appendix, we give the fitted parameter sets and the pole positions for Fit 1 and Fit 2 with $\Lambda=2.0-2.6$~GeV. 

\begin{table}[tb]
    \caption{\label{Tab_parametersFit1}Parameter sets from Fit 1 with different $\Lambda$ values. The uncertainties are the $1\sigma$ errors propagated from the statistical uncertainties of the data.}
    \renewcommand{\arraystretch}{1.2}
    \begin{tabular*}{\columnwidth}{@{\extracolsep\fill}lcccccccc}
    \hline\hline                                  
            $\Lambda~\rm{[GeV]}$
            & $\chi^2$/d.o.f. 
            & $a_1~\rm{[fm^2]}$
            & $a_2~\rm{[fm^2]}$
            & $b_1~\rm{[fm^2]}$
            & $b_2~\rm{[fm^2]}$
            & $c_1~\rm{[fm^2]}$
            & $\mathrm{Prp}^2$
            & $R_{n/p}$
            \\[3pt]     
    \hline
      2.0 &$0.54$ &$1.26_{-0.07}^{+0.09}$ &$-0.85_{-0.06}^{+0.06}$ &$1.57_{-0.07}^{+0.06}$ &$-0.83_{-0.06}^{+0.06}$ &$1.06_{-0.10}^{+0.07}$ &$0.06_{-0.02}^{+0.02}$ &$0.99_{-0.02}^{+0.01}$ \\[3pt]
      2.2 &$0.54$ &$1.28_{-0.06}^{+0.06}$ &$-1.69_{-0.05}^{+0.04}$ &$1.57_{-0.05}^{+0.06}$ &$-1.67_{-0.05}^{+0.04}$ &$1.13_{-0.08}^{+0.07}$ &$0.12_{-0.03}^{+0.04}$ &$0.99_{-0.01}^{+0.01}$ \\[3pt]
      2.4 &$0.54$ &$1.36_{-0.05}^{+0.05}$ &$-2.06_{-0.04}^{+0.03}$ &$1.63_{-0.05}^{+0.06}$ &$-2.04_{-0.03}^{+0.03}$ &$1.23_{-0.07}^{+0.07}$ &$0.18_{-0.05}^{+0.06}$ &$1.00_{-0.01}^{+0.00}$ \\[3pt]
      2.6 &$0.54$ &$2.15_{-0.04}^{+0.05}$ &$-1.10_{-0.03}^{+0.03}$ &$2.40_{-0.05}^{+0.04}$ &$-1.09_{-0.03}^{+0.03}$ &$2.02_{-0.08}^{+0.06}$ &$0.21_{-0.06}^{+0.07}$ &$0.99_{-0.01}^{+0.01}$ \\[3pt]
    \hline\hline
    \end{tabular*}
    \end{table}
    
    \begin{table}[tb]
    \caption{\label{Tab_parametersFit2}Parameter sets from Fit 2 with different $\Lambda$ values. The uncertainties are the $1\sigma$ errors propagated from the statistical uncertainties of the data.}
    \renewcommand{\arraystretch}{1.2}
    \begin{tabular*}{\columnwidth}{@{\extracolsep\fill}lcccccccc}
    \hline\hline                                  
            $\Lambda~\rm{[GeV]}$
            & $\chi^2$/d.o.f. 
            & $a_1~\rm{[fm^2]}$
            & $a_2~\rm{[fm^2]}$
            & $b_1~\rm{[fm^2]}$
            & $b_2~\rm{[fm^2]}$
            & $c_1~\rm{[fm^2]}$
            & $\mathrm{Prp}^2$
            & $R_{n/p}$
            \\[3pt]     
    \hline
      2.0 &$0.58$ &$-3.30_{-0.06}^{+0.07}$ &$-4.13_{-0.04}^{+0.04}$ &$-2.79_{-0.07}^{+0.07}$ &$-4.12_{-0.04}^{+0.03}$ &$-3.12_{-0.08}^{+0.08}$ &$0.54_{-0.09}^{+0.10}$ &$0.99_{-0.00}^{+0.00}$ \\[3pt]
      2.2 &$0.58$ &$-4.98_{-0.05}^{+0.07}$ &$-3.81_{-0.04}^{+0.04}$ &$-4.51_{-0.06}^{+0.06}$ &$-3.80_{-0.04}^{+0.04}$ &$-4.80_{-0.07}^{+0.06}$ &$0.88_{-0.19}^{+0.21}$ &$0.99_{-0.00}^{+0.00}$ \\[3pt]
      2.4 &$0.58$ &$-5.62_{-0.06}^{+0.06}$ &$-7.14_{-0.03}^{+0.03}$ &$-5.21_{-0.07}^{+0.05}$ &$-7.12_{-0.03}^{+0.03}$ &$-5.49_{-0.06}^{+0.06}$ &$2.30_{-0.45}^{+0.52}$ &$1.00_{-0.00}^{+0.00}$ \\[3pt]
      2.6 &$0.58$ &$-10.72_{-0.04}^{+0.04}$ &$-5.65_{-0.03}^{+0.03}$ &$-10.32_{-0.05}^{+0.04}$ &$-5.64_{-0.03}^{+0.03}$ &$-10.57_{-0.06}^{+0.05}$ &$4.68_{-1.04}^{+1.03}$ &$1.00_{-0.00}^{+0.00}$ \\[3pt]
    \hline\hline
    \end{tabular*}
\end{table}

The parameter sets for Fit 1 and Fit 2 with different values of $\Lambda$ are listed in Tables~\ref{Tab_parametersFit1} and \ref{Tab_parametersFit2}, respectively. For both fits, $a_2$ is constrained by unitarity to be negative, and $b_2$ as the imaginary part of the off-diagonal elements of the $\bm{V}_S$ matrix is also negative. $a_1$, $b_1$ and $c_1$ are positive for Fit 1 and negative for Fit 2. For both fits, the difference between $a_1$ and $c_1$ is small (less than 20\% in Fit 1 and less than 5\% in Fit 2), meaning that the isospin breaking effects are subleading. The short distance production sources for the $p\bar{p}$ and $n\bar{n}$ are almost identical, as their ratio $R_{n/p}\simeq 1$ in both fits. 

\begin{figure}[tb]
{\includegraphics[width=0.496\textwidth]{figs/poles_2c_fit1_L2_5p.pdf}}
{\includegraphics[width=0.496\textwidth]{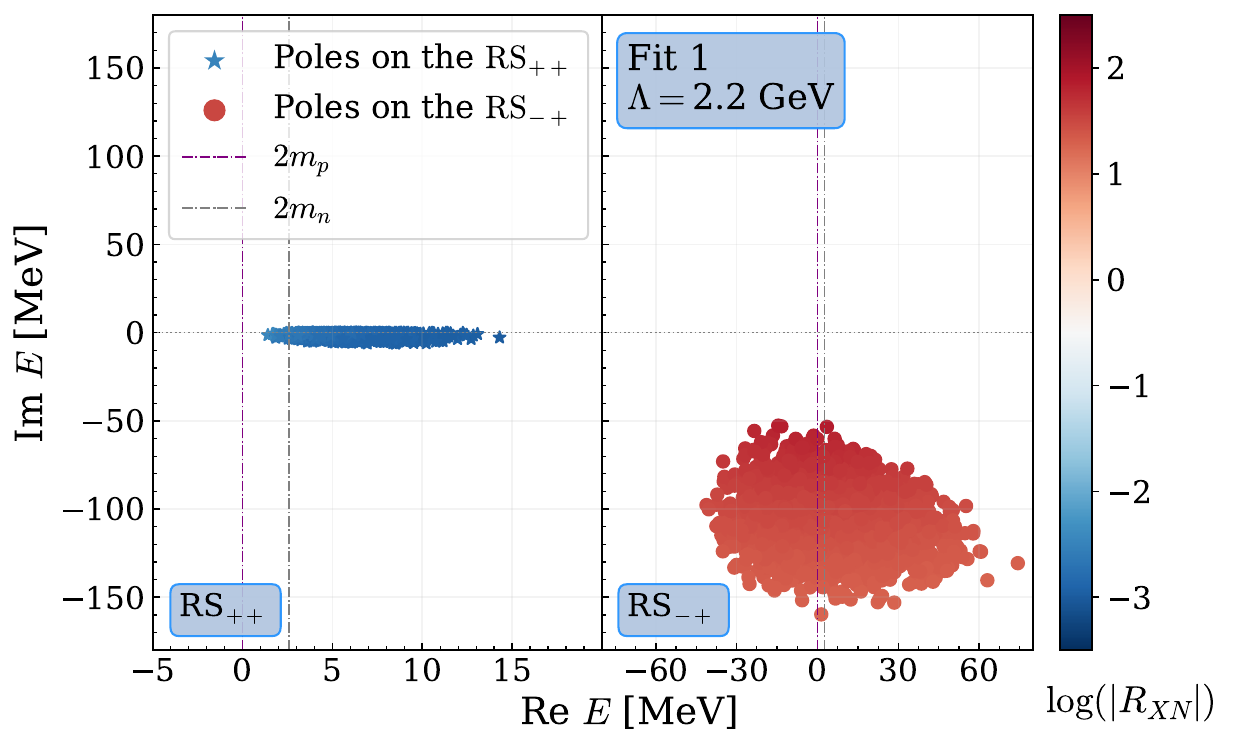}}
{\includegraphics[width=0.496\textwidth]{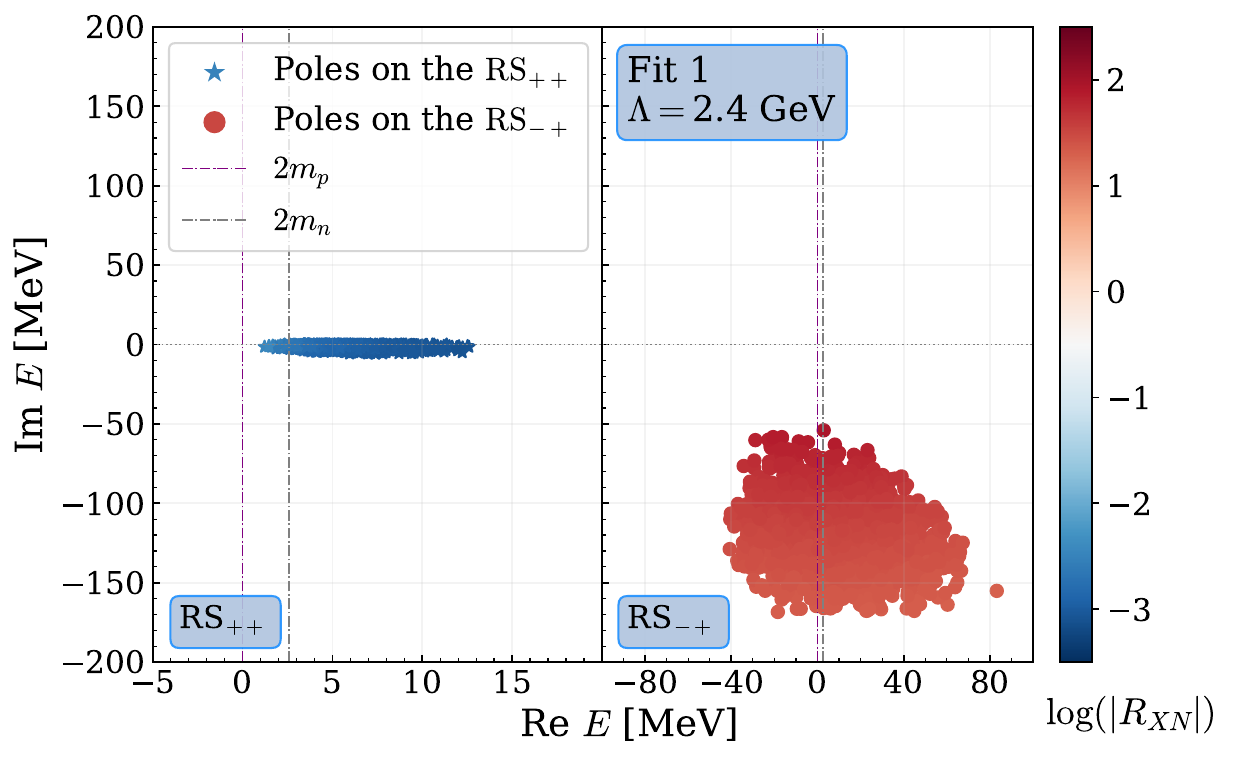}}
{\includegraphics[width=0.496\textwidth]{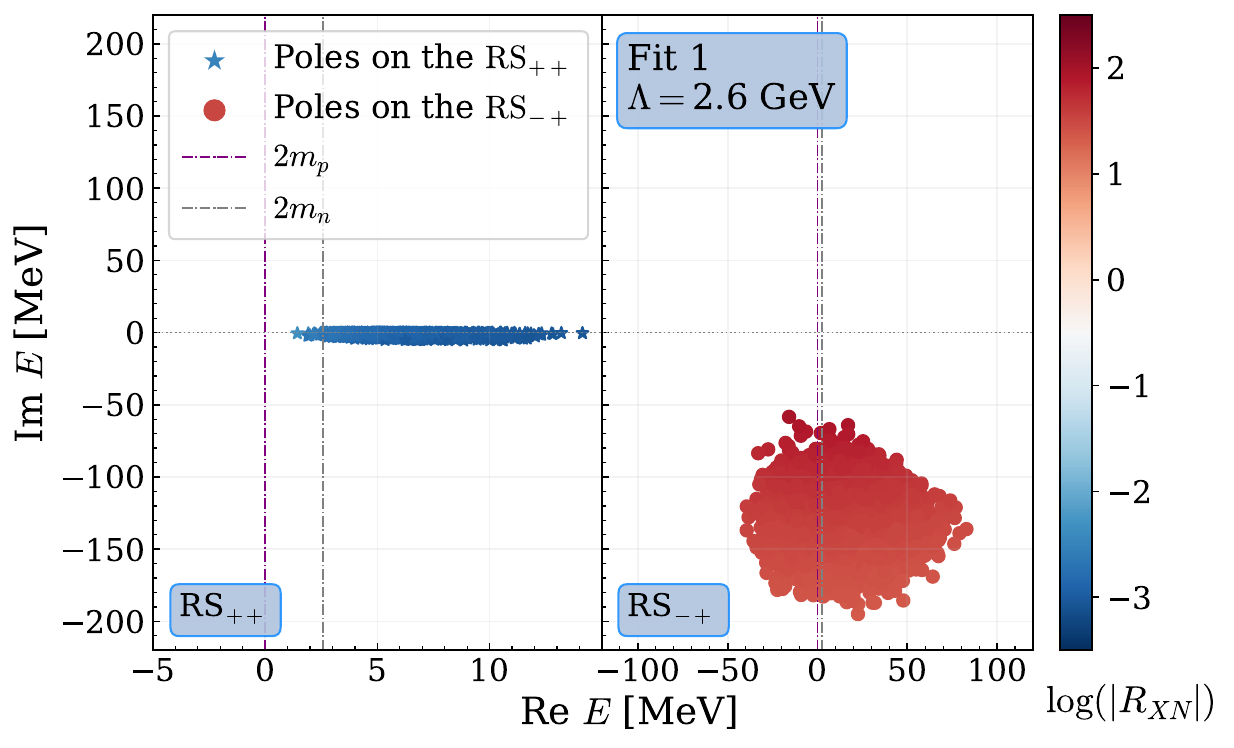}}
\caption{Pole positions from Fit 1 with different $\Lambda$ values using parameters in the 1$\sigma$ error range. The color coding is the same as in Fig.~\ref{Fig_poleFit_L2}.}
\label{Fig_poleFit1}
\end{figure}

\begin{figure}[tb]
{\includegraphics[width=0.496\textwidth]{figs/poles_2c_fit2_L2_5p.pdf}}
{\includegraphics[width=0.496\textwidth]{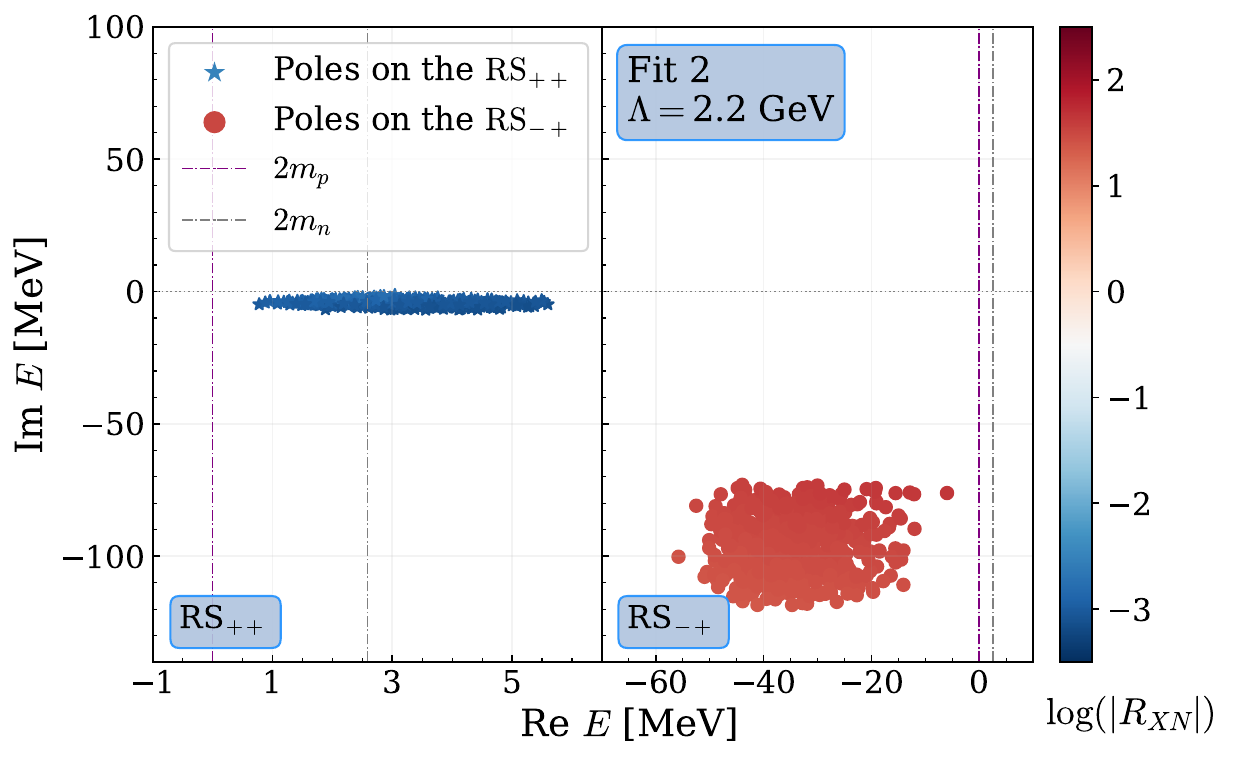}}
{\includegraphics[width=0.496\textwidth]{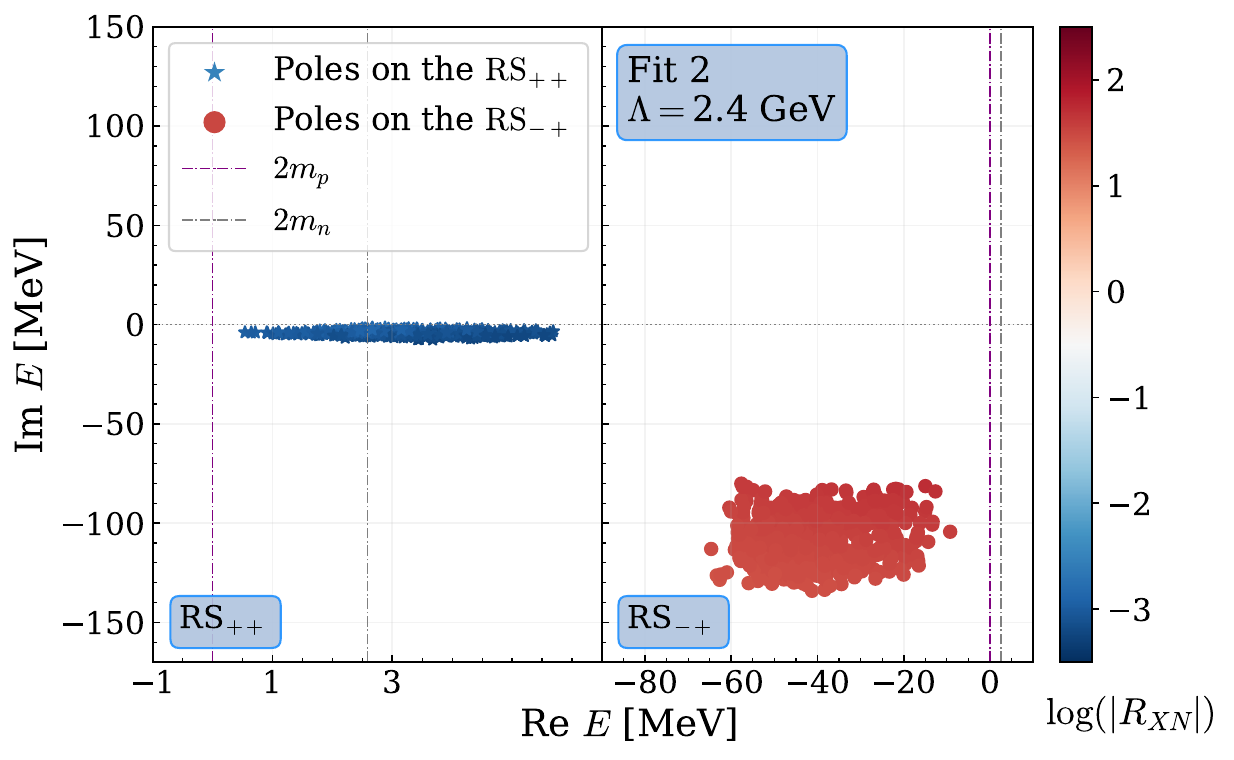}}
{\includegraphics[width=0.496\textwidth]{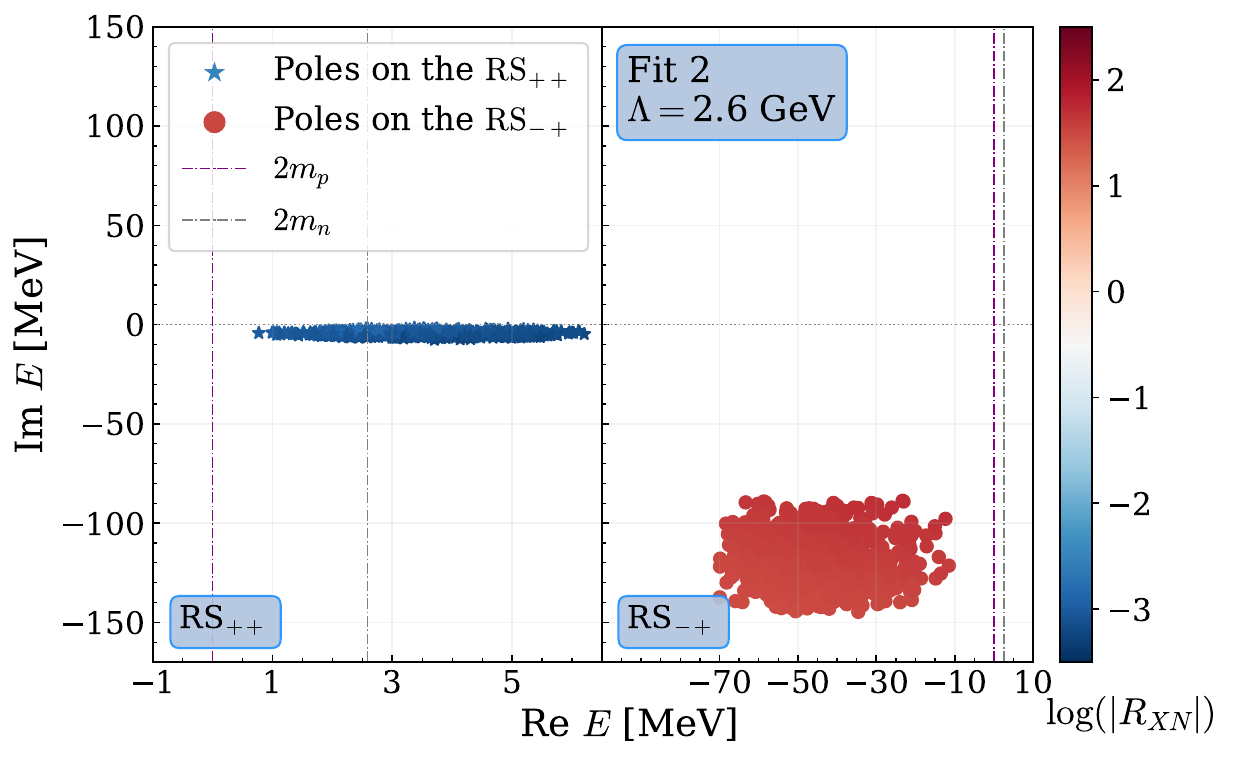}}
\caption{Pole positions from Fit 2 with different $\Lambda$ values using parameters in the 1$\sigma$ error range. The color coding is the same as in Fig.~\ref{Fig_poleFit_L2}.}
\label{Fig_poleFit2}
\end{figure}

\begin{figure}[tb]
{\includegraphics[width=0.496\textwidth]{figs/poles_2c_L2_err_fixa11eff.pdf}}
{\includegraphics[width=0.496\textwidth]{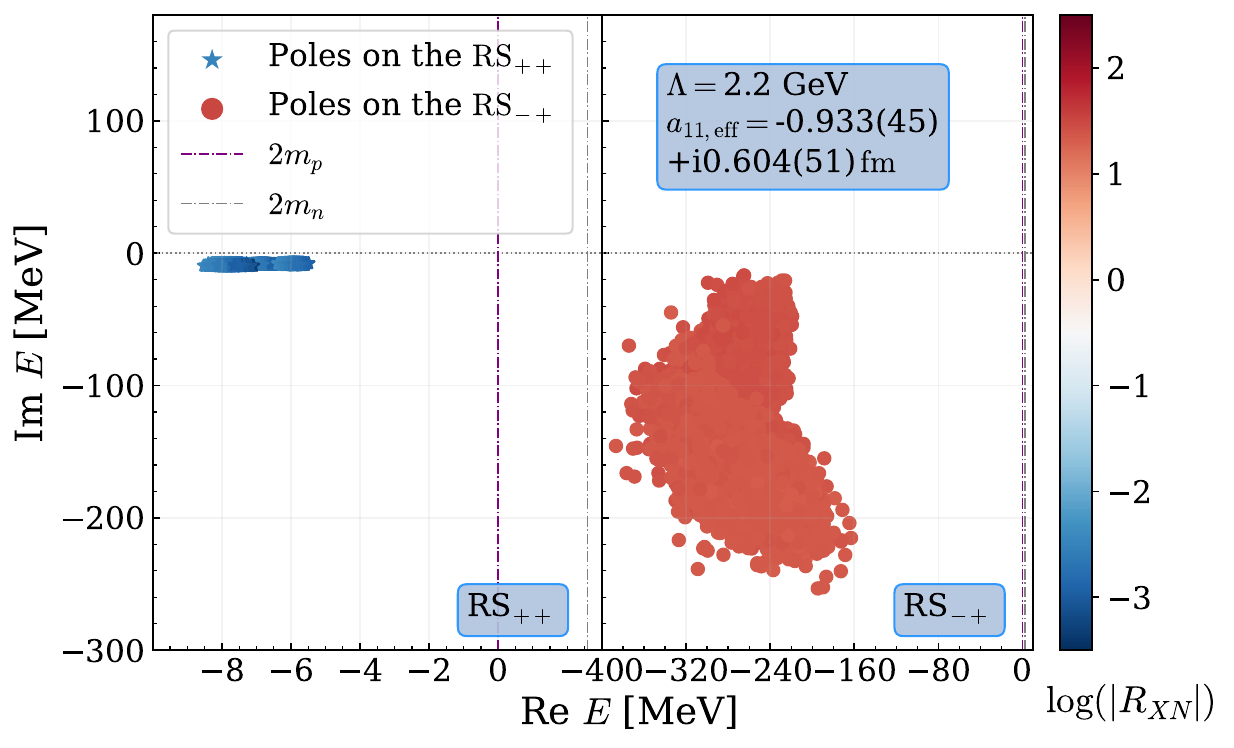}}
{\includegraphics[width=0.496\textwidth]{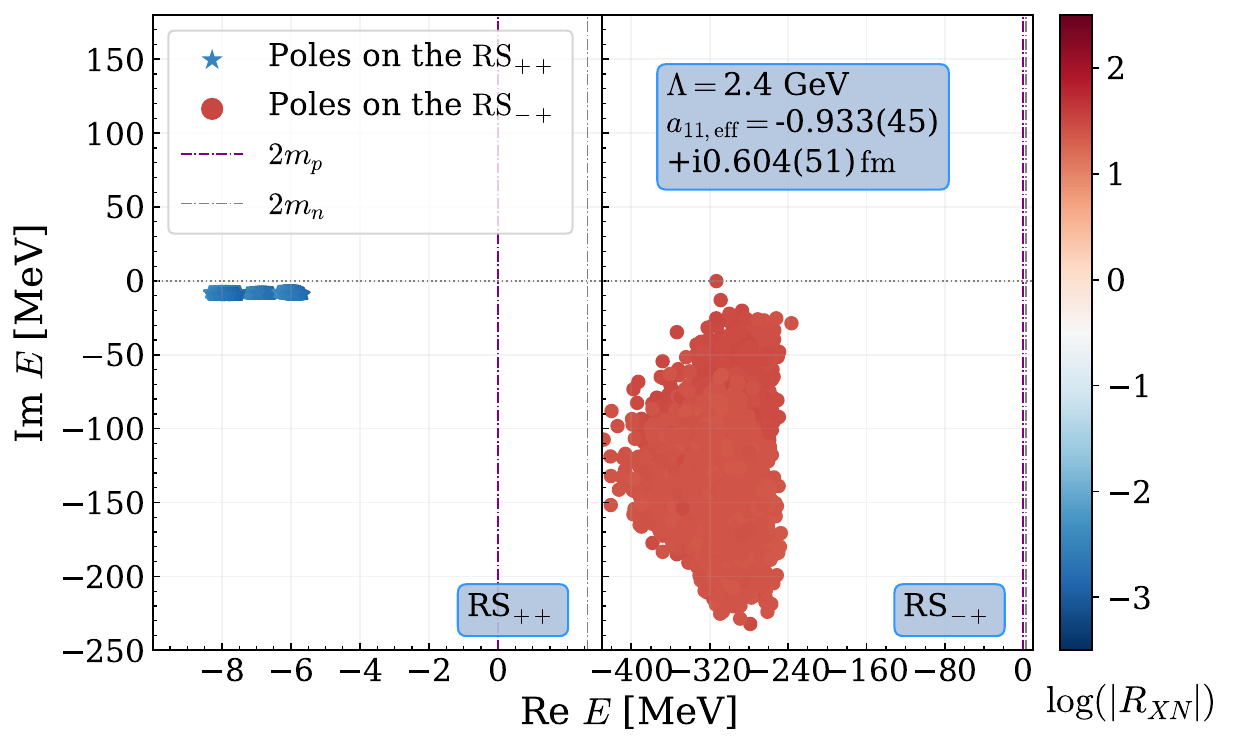}}
{\includegraphics[width=0.496\textwidth]{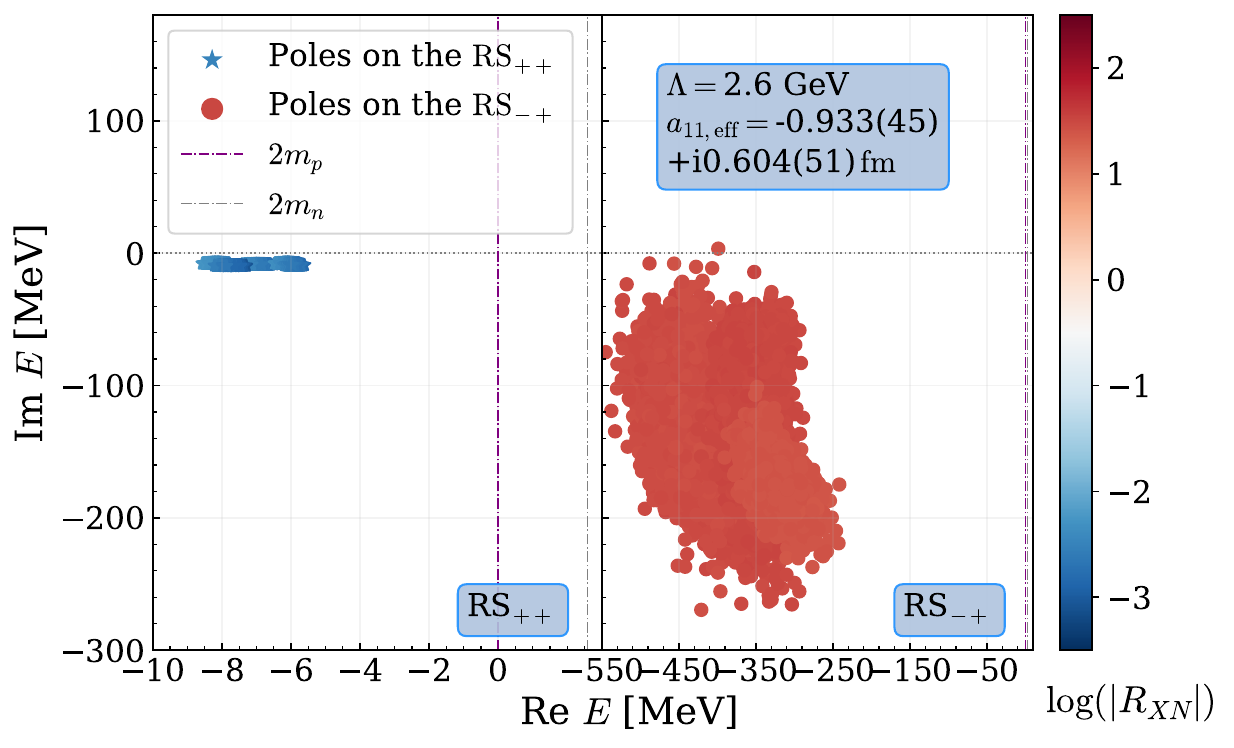}}
\caption{Pole positions from fit with $a_{11,\rm{eff}}=-0.933(45)+i0.604(51)~\rm{fm}$~\cite{Gotta:2004rq,Carbonell:2023onq} for different $\Lambda$ values. The uncertainties are same as Fig.~\ref{Fig_poles_fixeda11}, and the color coding is the same as in Fig.~\ref{Fig_poleFit_L2}.}
\label{Fig_poleFit_fixeda11}
\end{figure}

The near-threshold pole positions of the $\bm{T}_{SC}$ matrix with different $\Lambda$ values for Fit 1 and Fit 2 are shown in Figs.~\ref{Fig_poleFit1} and \ref{Fig_poleFit2}, respectively, and the pole positions for the fit with fixed $a_{11,\,\text{eff}}$ are shown in Fig.~\ref{Fig_poleFit_fixeda11}, which are farther from threshold than those in Fits 1 and 2. Their features have been discussed in the main text.

\clearpage
\bibliography{ppbar.bib}
\end{document}